\documentclass[preprint,3p,12pt]{elsarticle}
\usepackage{amssymb,amsmath,latexsym,braket,fancyref,hyphenat,float,tablefootnote}

\numberwithin{equation}{section}

\biboptions{numbers,sort&compress}

\begin{document}

\begin{flushright}
MITP/17-010\\ 
MAN/HEP/2017/01\\
November 2017\\[-1cm]
${}$
\end{flushright}

\vspace{0.5em}

\begin{frontmatter}

\title{
{\bf Frame Covariant Nonminimal Multifield Inflation}
}

\author[man]{Sotirios Karamitsos} 
\ead{sotirios.karamitsos@manchester.ac.uk}
\author[man,mitp]{Apostolos Pilaftsis}
\ead{apostolos.pilaftsis@manchester.ac.uk}

\address[man]{Consortium for Fundamental Physics, School of Physics and Astronomy,\\
 University of Manchester, Manchester M13 9PL, United Kingdom}
\address[mitp]{PRISMA Cluster of Excellence \& Mainz Institute for Theoretical Physics,\\
Johannes Gutenberg University, 55099 Mainz, Germany}

\begin{abstract} 
{\small 
We introduce a frame-covariant formalism for inflation of scalar-curvature theories by adopting a differential geometric approach which treats the scalar fields as coordinates living on a field-space manifold. This ensures that our description of inflation is \emph{both} conformally and reparameterization covariant. Our formulation gives rise to extensions of the usual Hubble and potential slow-roll parameters to generalized fully frame-covariant forms, which allow us to provide manifestly frame-invariant predictions for cosmological observables, such as the tensor-to-scalar ratio~$r$, the spectral\- indices $n_{\cal R}$ and $n_T$, their runnings $\alpha_{\cal R}$ and $\alpha_T$, the non-Gaussianity parameter~$f_{NL}$, and the isocurvature fraction $\beta_{\rm iso}$.  We examine the role of the field space curvature in the generation\- and transfer of isocurvature modes, and we investigate the effect of boundary conditions for the scalar fields at the end of inflation on the observable inflationary quantities.  We explore the stability\- of the trajectories with respect to the boundary conditions by using a suitable sensitivity parameter. To illustrate our approach, we first analyze a simple minimal two-field scenario before studying a more realistic nonminimal model inspired by Higgs inflation. We find that isocurvature effects are greatly enhanced in the latter scenario and must be taken into account for certain values in the parameter space such that the model is properly normalized to the observed scalar power spectrum~$P_{\cal R}$. Finally, we outline how our frame-covariant approach may be extended beyond the tree-level approximation through the Vilkovisky--De Witt formalism, which we generalize  to take into account conformal transformations, thereby leading to a fully frame-invariant effective action at the one-loop level.

\medskip
\noindent {\sc Keywords}: Inflation, Scalar-Curvature Theories, Multifield Models, Frame Covariance }
\end{abstract}

\end{frontmatter}

\newpage

\tableofcontents

\newpage

\section{Introduction}
\label{intro}

The framework of inflation, originally conceived as a way to resolve the flatness and horizon problems, has been extremely successful in explaining the origin of cosmological perturbations~\cite{Mukhanov:1990me, Ratra:1987rm,Lyth:1998xn}. Numerous inflationary models with various theoretical motivations from particle physics, including supergravity and axion inflation, have been proposed so far~\mbox{\cite{Mazumdar:2010sa,Yamaguchi:2011kg,Pajer:2013fsa}}. The simplest and most thoroughly studied paradigm involves a single scalar field that drives the early exponential expansion of the Universe. One of the most straightforward extensions of such theories is multifield inflation, a class of models featuring more than one scalar field contributing to the inflationary expansion of the Universe~\cite{Starobinsky:2001xq, Senatore:2010wk}. While the current cosmological data are well described by single-field inflation \cite{Ade:2013zuv,Ade:2015lrj}, models of multifield inflation are of great theoretical interest, as they provide new predictions that could be tested by future observations~\cite{Kaiser:2013sna,Schutz:2013fua}.

When dealing with nonminimal models, one has to necessarily contend with the so-called \emph{frame problem}. This problem pertains to the question of whether inflationary models related by a \emph{frame transformation}, namely a local rescaling of the metric followed by a field reparameterization, are physically equivalent \cite{Dicke:1961gz,faraoni98}. While the consensus is that no physical difference exists between the frames at the classical level if care is taken to appropriately transform all quantities~\cite{ Flanagan:2004bz,Chiba:2013mha,Postma:2014vaa,Burns:2016ric,Jarv:2016sow}, the situation beyond the tree level is far from clear~\cite{Steinwachs:2013tr,Kamenshchik:2014waa,Domenech:2015qoa}. There is no widely accepted procedure which ensures that radiative corrections to inflationary models are themselves frame-independent, and as such, there has been much discussion as to whether a particular frame is the ``physical'' one, or whether choosing a frame is a mere mathematical convenience.

In this article, we extend the covariant formalism for multifield inflation developed in~\cite{vanTent:2003mn,Achucarro:2010da} and extended in~\cite{Gong:2011uw} beyond the linear level to the more general framework of {\em nonmininal} multifield scalar-curvature theories. We pay particular attention to conformal transformations and promoting reparameterization covariance to \emph{full} frame covariance. The cosmological observables of interest to us are: the tensor-to-scalar ratio $r$, the spectral indices $n_{\cal R}$ and~$n_T$, their runnings $\alpha_{\cal R}$ and $\alpha_T$, the non-Gaussianity parameter~$f_{NL}$, and the isocurvature fraction $\beta_{\rm iso}$. Our formalism addresses the frame problem at the classical level in a way that allows for its direct extension beyond the tree level, where quantum loop effects are taken into account. In particular, to go beyond the Born approximation, we employ the Vilkovisky--De Witt formalism~\cite{Vilkovisky:1984st,Rebhan:1986wp,DeWitt} with a conformal extension in order to derive a fully frame-invariant effective action at the one-loop level.  We argue that the so-derived Vilkovisky--De Witt effective action can be used to compute frame-invariant radiative corrections to cosmological observables, giving rise to the same predictions for both the Jordan and Einstein frames.

The outline of this paper is as follows: in Section~\ref{sec:frametrans}, we present the classical action for the class of theories that we will be studying, specified by three model functions: (i)~the nonminimal coupling $f(\varphi)$, (ii)~the multifield wavefunction $k_{AB}(\varphi)$, and (iii)~the scalar potential~$V(\varphi)$, where $\varphi$ collectively stands for all the scalar fields. By considering their properties under conformal transformations and field reparametrizations, we show that the functional form of the classical action remains invariant under frame transformations. In this way, we introduce the concept of \emph{frame covariance} in inflation by defining frame-covariant extensions to well-known cosmological quantities such as the Hubble parameter and the comoving density and pressure, enabling us to write the inflationary equations of motion in a manifestly frame-covariant manner. 

In Section~\ref{quantpert}, we define the concept of a \emph{field space}, treating the scalar fields $\varphi$ as coordinates on a manifold where inflationary trajectories reside. We distinguish between curvature and isocurvature perturbations and make a connection between the primordial perturbations and the observable power spectra. We briefly discuss the super-horizon evolution of perturbations, which becomes relevant in the presence of isocurvature perturbations in multifield theories. We extend the usual definitions of the slow-roll hierarchy and Hubble slow-roll parameters to frame-covariant forms.  Specializing to two-field models, we examine the effect of the entropy transfer by deriving approximate analytical results for the transfer functions and we study the effects that a curved field space might have on the amplification and the transfer of isocurvature modes. Finally, we investigate the effect of boundary conditions for the scalar fields at the end of inflation on the observable inflationary quantities and the stability of inflationary trajectories.

In Section~\ref{specmod}, we apply our formalism to two specific models:  (i) a simple minimal two-field model with a light scalar field and a small quartic coupling, and (ii) a nonminimal model inspired by Higgs inflation \cite{Bezrukov:2007ep}. We parameterize the boundary conditions on the end-of-inflation isochrone curve, and we use the normalization of the observed scalar power spectrum~$P_{\cal R}$ to select a valid inflationary trajectory. Noting that the minimal model is not observationally viable, we modify it by including a nonminimal coupling $\xi$ between one of the light scalar fields and the Ricci scalar~$R$. Upon choosing a nominal value for~$\xi$, we find that isocurvature effects are significant  in obtaining predictions for inflationary observables that are compatible with cosmological observations. Finally, we outline how $F(\varphi,R)$ theories can be incorporated in our formalism, written in terms of an equivalent multifield inflation model by using the method of Lagrange multipliers.

In Section~\ref{radcorr}, we present the Vilkovisky--De Witt formalism, which was originally developed in order to solve the apparent non-uniqueness problem of the effective action under field reparameterizations. Under the assumption that gravitational corrections can be neglected, we outline the fundamentals of the Vilkovisky--De Witt formalism and how it can be applied to theories of multifield inflation. In analogy to our conformally covariant extension of inflation at the tree level, we extend the Vilkovisky--De Witt formalism to take into consideration conformal transformations, which we expect to be essential in future computations for fully frame-covariant radiative corrections to inflationary quantities. Finally, Section~\ref{conclusion} summarizes our findings and presents  possible future directions for further research.

\section{Frame Transformations and Classical Dynamics in Multifield Inflation}
\label{sec:frametrans}

In this section, we specify the class of models that we will be studying by defining their classical action and examining their properties under frame transformations. Hence, we introduce the concept of \emph{frame covariance} in inflation by defining extensions to well-known cosmological quantities, such as the Hubble parameter $H$, the comoving energy density~$\rho$ and pressure~$P$.  In this way, we are able to recast the  equations of motion pertinent to inflation in a manifestly frame-covariant manner. 

The class of models of interest to us may be described by the following multifield scalar-curvature action:
 \begin{align}
\label{actionJ}
S\ \equiv\ \int  d^4 x\,  \sqrt{-g}  \, \left[-\frac{f(\varphi)}{2} R + \frac{k_{AB}(\varphi)}{2} g^{\mu\nu }(\nabla_\mu \varphi^A) (\nabla_\nu \varphi^B) - V(\varphi)  \right],
\end{align}
where $g_{\mu\nu}$ is the spacetime metric whose determinant is denoted by $g \equiv \det g_{\mu\nu}$,  $f(\varphi)$ is the nonminimal coupling to the Ricci scalar $R$, $k_{AB}$ is the multifield wavefunction, and~$V(\varphi)$ is the scalar potential. These model parameters are in general functions of $\varphi$ which, without any indices, collectively stands for all the scalar fields $\varphi^A$.  In this notation, uppercase indices~$A, B, \ldots$  run over the different fields. Moreover, we assume that the energy density of the scalar fields dominates the action during inflation, and so the matter sector can be neglected. Finally, our convention for the Minkowski flat limit of $g_{\mu\nu}$ is~$\eta_{\mu\nu} = {\rm diag}\, (+1,-1,-1,-1)$.

The  action~$S$ given in~\eqref{actionJ} is said to be defined in the \emph{Jordan frame}, in which 
the nonminimal coupling $f(\varphi)$ replaces the squared Planck mass~$M^2_P$ of minimal models. The coupling~$f(\varphi)$ can be modified by a {\em frame transformation}, which consists of a conformal transformation
\begin{equation}
\label{conftrans}
\begin{aligned}
g_{\mu \nu}\ \mapsto\ \tilde g_{\mu\nu} \hspace{0.21em}  &=\   \Omega^2  \, g_{\mu\nu}\;,
\\
\varphi^A\  \mapsto\ \widetilde \varphi^A \ &=\   \Omega^{-1}\, \varphi^A \;,
\end{aligned}
\end{equation}
followed by a field reparameterization 
 \begin{align}
 \label{infrep}
\hspace{0.2em}
\varphi^A\ \mapsto\  \varphi^{\widetilde A} \ =  \ \varphi^{\widetilde A} (  \varphi)  \;.
\end{align}
In the above, an index with a tilde corresponds to a transformation to a new set of fields, in analogy to a diffeomorphism between two sets of coordinates. For a general conformal transformation, the conformal factor $\Omega=\Omega(x)$ is a function of spacetime, but we restrict our attention to conformal factors $\Omega=\Omega(\varphi)$ that depend on $x$ only through the scalar fields~$\varphi^A = \varphi^A(x)$.  The field reparameterization \eqref{infrep} has an associated Jacobian given by
 \begin{align}
 \label{reparam}
\frac{d  \varphi^{\widetilde A}}{d\varphi^B}\ &\equiv \ J^{\widetilde  A}_{\ B} (\varphi) \;.
\end{align}
Thus, the full frame transformation may be written as
\begin{equation}
\label{frametrans}
\begin{aligned}
g_{\mu\nu}\ \mapsto\ \tilde g_{\mu\nu} \hspace{0.21em} &=\   \Omega^2  \;  g_{\mu\nu}\;,
\\
\varphi^A\ \mapsto\ \widetilde \varphi^{\widetilde A} \ &=\   \Omega^{-1}\, 
\varphi^{\widetilde A}(\varphi )\;.
\end{aligned}
\end{equation}
The frame transformation \eqref{frametrans} induces a field reparameterization: $\varphi^A \mapsto \widetilde \varphi^{\widetilde A} =\widetilde \varphi^{\widetilde A}(\varphi)$, which differs from \eqref{infrep}, and has the following associated Jacobian:
 \begin{align}
 \label{jacobian}
\frac{d \widetilde \varphi^{\widetilde A}}{d\varphi^B}\ &= \  \Omega^{-1} \left[ J^{\widetilde A}_{\ B} -  \varphi^{\widetilde A} (\varphi ) \ln \Omega_{,B}\right]\; \
\nonumber \\
&\equiv\  \Omega^{-1} K^{ \widetilde A}_{\ B}(\varphi) \;,
\end{align}
where $\Omega_{,A} \equiv \partial \Omega/\partial\varphi^A$ and 
$K^{\tilde A}_{\ B} (\varphi)$ has been defined in such a way that it does not contain the prefactor~$\Omega^{-1}$. 

It is possible to choose $\Omega$, so that $f(\varphi) = M^2_P$, where $M_P$ is the reduced Planck mass. In this case, it is said that the action~$S$ is defined in the \emph{Einstein frame}. Hence, the \emph{frame problem} relates to the question of whether different frames describe the same physics. Whilst this problem is resolved at the classical level, it remains an open question beyond the tree level. Our aim is to build upon the classical treatment, which will assist us in examining the uniqueness of quantum loop effects on inflation in different frames in Section \ref{radcorr}. For this reason, we make no \emph{a priori} assumptions about frame invariance, even at the classical level.

We begin by characterizing a quantity $X^{A_1 A_2 \ldots A_p}_{B_1 B_2 \ldots B_q}$ to be frame covariant if it obeys the following two transformation properties under a frame transformation \eqref{frametrans}:
\begin{align}
 \label{confdef}
\widetilde X^{  A_1   A_2 \ldots A_p}_{B_1 B_2 \ldots A_q}   \ &=\ \Omega^{-w_X }  X^{  A_1   A_2 \ldots A_p}_{B_1 B_2 \ldots B_q}\;  ,
\\
 \label{repfdef}
  X^{\widetilde A_1 \widetilde A_2 \ldots \widetilde A_p}_{\widetilde B_1 \widetilde B_2 \ldots \widetilde B_q} \ &=\ \Omega^{-(p-q)} (K^{\widetilde A_1}_{\ A_1} K^{\widetilde A_2}_{\ A_2} \ldots K^{\widetilde A_p}_{\ A_p} ) \ X^{  A_1   A_2 \ldots A_p}_{B_1 B_2 \ldots B_q}\  (K^{ B_1}_{\ \widetilde B_1}K^{ B_2} _{\ \widetilde B_2}\ldots K^{ B_q}_{\ \widetilde B_q})\; ,
\end{align}
where $w_X$ is the \emph{conformal weight} of the quantity $X^{A_1 A_2 \ldots A_p}_{B_1 B_2 \ldots B_q}$, which does not depend on its number of indices. In order to avoid notational clutter, we suppress arguments of $\varphi$ here and in the following. The first property \eqref{confdef} corresponds to the conformal transformation of the quantity itself, whereas the second one~\eqref{repfdef} is due to the diffeomorphism encoded {\em via} the Jacobian \eqref{jacobian}. The frame transformation \eqref{frametrans} applied to $X^{A_1 A_2 \ldots A_p}_{B_1 B_2 \ldots B_q}$ combines the two transformation properties \eqref{confdef} and \eqref{repfdef} as follows:
\begin{align}
\label{covdef}
\widetilde  X^{\widetilde A_1 \widetilde A_2 \ldots \widetilde A_p}_{\widetilde B_1 \widetilde B_2 \ldots \widetilde B_q} \ &=\ \Omega^{-d_X}  (K^{\widetilde A_1}_{\ A_1} K^{\widetilde A_2}_{\ A_2} \ldots K^{\widetilde A_p}_{\ A_p} )  
\ X^{  A_1   A_2 \ldots A_p}_{B_1 B_2 \ldots B_q} \ 
(K^{ B_1}_{\ \widetilde B_1}K^{ B_2} _{\ \widetilde B_2}\ldots K^{ B_q}_{\ \widetilde B_q})\;,
\end{align}
where $d_X$ denotes the \emph{scaling dimension} of $X$, given by
\begin{align}
\label{scaldimrel}
d_X\ =\ w_X + p - q\; .
\end{align}
We thus see that we may assign to $X$ a scaling dimension of $+1$ for every contravariant index and $-1$ for every covariant one, in addition to its weight  $w_X$ induced by a conformal transformation. For example, with the above convention, the spacetime metric $g_{\mu\nu}$ (which carries no field indices) has conformal weight and scaling dimension both equal to~$-2$.

In order to determine whether the action~$S$ in~\eqref{actionJ} is frame invariant, it is instructive to study its response under a general frame transformation. In this case, the Ricci scalar~$R$ transforms to
\begin{align}
\widetilde R\ =\ \Omega^{-2}R\: -\:  6 \Omega^{-3}\: (\nabla^\mu\nabla_\mu \Omega)\; .
\end{align}
Likewise, the model functions transform according to the following rules~\cite{Flanagan:2004bz,Jarv:2014hma}:
\begin{equation}
\label{transrules}
\begin{aligned}
\tilde f\  &=\ \Omega^{-2}\,f \;, \\ 
{\tilde k}_{ \widetilde A  \widetilde B}\ &=\   \left(k_{AB}   - 6   f \Omega^{-2 }  \Omega_{,A} \Omega_{,B} 
  + 3 \Omega^{-1 }   f_{,A}\Omega_{,B} 
  +3\Omega^{-1} \Omega_{,A}  f_{,B} \right) K^A_{\  \widetilde A} K^B_{\ \widetilde   B}\;,\\
 \widetilde V\  &=\ \Omega^{-4}\,V\;.
\end{aligned}
\end{equation}
Given~\eqref{covdef}, we can see from~\eqref{transrules} that the model functions $f $ and $V $ are frame covariant, with scaling dimension $2$ and $4$, respectively. Instead, $k_{AB}$ is not a  frame-covariant quantity. However, these transformation rules may be used to show that the action~$S$ is form invariant, i.e.
\begin{align}
S[g_{\mu\nu}, \varphi, f , k_{AB},V]\  =\  S[\tilde g_{\mu\nu}, \widetilde \varphi, \tilde f , \tilde k_{AB},\widetilde V]\; .
\end{align}
This equivalence is the starting point for our formalism, since it ensures that any results derived for one frame must apply to any other frame. It also indicates that this frame covariance must be reflected at the level of the equations of motion. Therefore, our goal in the remainder of the section is to derive the equations of motion for a general multiscalar-curvature theory, and show that they can be rewritten in a manifestly frame-covariant form.

In order to derive the equations of motion, we vary the action~$S$ with respect to the scalar fields $\varphi^A$ and the metric $g_{\mu\nu}$. Varying $S$ with respect to~$\varphi^A$ yields
 \begin{align}
 \label{infeq}
 k_{AB} \nabla_\mu \nabla^\mu \varphi^B
&+ \left(   \frac{k_{AB,C}}{2}  
+\frac{ k_{CA,B}}{2}  
- \frac{ k_{BC,A}}{2}\right)\, (\nabla_\mu \varphi^B) (\nabla^\mu \varphi^C)  
+ \frac{f_{,A}}{2}R 
+V_{,A}\  =\ 0\; .
\end{align}
Moreover, varying $S$ with respect to the metric~$g_{\mu\nu}$ gives rise to the Einstein equation
 \begin{align}
 \label{einsteineq}
G_{\mu\nu}\ =\ M_P^{-2} T^\text{(NM)}_{\mu\nu}\; ,
\end{align}
where the nonminimal energy-momentum tensor $T^\text{(NM)}_{\mu\nu}$ is modified due to the presence of the nonminimal coupling $f $. Its analytic form is given by
\begin{align}
   \label{effset}
M_P^{-2} T^\text{(NM)}_{\mu\nu}\ =\ 
\frac{T_{\mu \nu}}{f}  - \frac{f_{,AB}}{f} & (\nabla_\rho  \varphi^A)(\nabla^\rho  \varphi^B) g_{\mu\nu}  
-  \frac{f_{,A}}{f} (\nabla^2 \varphi^A) g_{\mu\nu}  
\nonumber\\
&
+ \frac{f_{,A}}{f}\,(\nabla_\mu \nabla_\nu \varphi^A) 
+  \frac{f_{,AB}}{f}\,(\nabla_\mu \varphi^A) ( \nabla_\nu \varphi^B)\;,
\end{align}
where
\begin{align}
T_{\mu\nu}\ &=\   
 k_{AB}   (\nabla_\mu \varphi^A) (\nabla_\nu \varphi^B)  
- \frac{k_{AB} }{2}\,(\nabla_\rho \varphi^A) (\nabla^\rho \varphi^B) g_{\mu\nu}  
+ V g_{\mu\nu}
\end{align}
is the standard energy-momentum tensor.  Equations~\eqref{infeq} and~\eqref{einsteineq} are the equations of motion that govern the evolution of the scalar fields given a general curved background. These equations are of particular cosmological interest when the scalar fields $\varphi$ are spatially homogeneous, i.e.~$\varphi^A =\varphi^A (\tau)$, and when the metric $g_{\mu\nu}$ takes on the well-known  Friedmann--Robertson--Walker~(FRW) form:
  \begin{align}
ds^2\ =\ g_{\mu\nu}dx^\mu dx^\nu\ =\ N_L^2 d\tau^2 - a ^2 \left( \frac{dr^2}{1-kr^2}+r^2 d\Omega \right).
 \end{align}
 Here, $a = a(\tau)$ is the \emph{scale factor}, $N_L = N_L(\tau)$ is the \emph{lapse function}, $d\Omega$ is the three-dimensional solid angle element, and $k$ is the curvature which is set to zero in the following.
 
Given the assumptions of homogeneity of the fields and the FRW form for the metric, we may eliminate $R$ from the scalar field equation~\eqref{infeq} by taking the trace of~\eqref{einsteineq}, resulting in
  \begin{align}
 \label{infeqFRW}
0\ =&\ \left( k_{AB}  + \frac{3   f_{,A} f_{,B}}{2 f } \right) \left[{\ddot \varphi}^B +   \big(3H + H_L\big)     {\dot \varphi}^B\right]
 \nonumber \\
&\ +\frac{1}{2}  \left[   k_{BA,C}   
+  k_{AC,B} 
-  k_{BC,A} 
+  \frac{f_{,A}}{ f } \big( k_{BC}    + 3  f_{,BC}  \big)
 \right] {\dot \varphi}^B  {\dot \varphi}^C
+N_L^2 f^2  \left(\frac{V}{f^2}\right)_{,A}  ,
 \end{align}
where the \emph{Hubble parameter} $H$ and the \emph{lapse rate} $H_L$ are given by
   \begin{align}
H\ \equiv\ \frac{\dot a}{   a}\;, \qquad H_L\: \equiv\ \frac{\dot N_L}{N_L }\; ,
 \end{align}
and the overdot denotes differentiation with respect to the coordinate~$\tau$. In addition, the Friedmann and acceleration equations may be derived via the temporal and spatial components of~\eqref{einsteineq}:
\begin{align}
\label{friedeq}
  H^2\ &=\  \frac{1}{3f} \left( \frac{k_{AB}\dot \varphi^A \dot \varphi^B}{2 } + N_L^2 V \right) - \frac{  H \dot f}{f}\;,
\\
\label{acceleq}
 \dot H -H_L H\  &=\ - \frac{1}{2f}  \left( \frac{k_{AB}\dot \varphi^A \dot \varphi^B}{2} \right) + \frac{ H \dot f}{2f} - \frac{\ddot f}{2f}\;.
\end{align}
These equations of motion appear in the literature in various forms. Specifically, for minimal inflation models, $f$ is set to $M_P^2$, whereas for single-field inflation the fields $\varphi^A$ are replaced by $\varphi$. Note that the cosmological equations~\eqref{infeqFRW},~\eqref{friedeq}, and~\eqref{acceleq} are not manifestly frame invariant as written above. Setting $N_L$ equal to unity is a common procedure in the literature as it makes calculations easier when working in a given frame. For that reason, it is crucial to work with a generic lapse function $N_L$ when working in the context of frame transformations if we do not wish for the coordinates to transform, thereby fully encoding the frame transformation in a transformation of the metric~$g_{\mu\nu}$.

With the aim to write the equations of motion in a manifestly covariant manner,  we define, in terms of the model functions, the basic frame-covariant quantities~\cite{Kaiser:2010ps,Hohmann:2016yfd}
\begin{align}
   \label{eq:GAB}
G_{AB}\  &\equiv\ \frac{k_{AB}}{f} + \frac{3}{2} \frac{f_{,A} f_{,B}}{f^2}\; ,
\qquad
U \ \equiv\ \frac{V}{f^2}\; .
\end{align}
Under \eqref{covdef}, these quantities transform as
\begin{align}
\widetilde G_{\widetilde A \widetilde B}\  &= \Omega^{2} \ G_{AB}\, K^A_{\ \widetilde A}K^B_{\ \widetilde B}  \; ,
\qquad
 \widetilde U \ =\ U \;,
\end{align}
because $w_G = 0$, $d_G = -2$ and $w_U = d_U = 0$.

Our next task is to define a derivative that respects the covariant properties of the quantities on which it acts. From \eqref{covdef}, we expect a proper frame-covariant field derivative to satisfy the following transformation property:
\begin{align}
\label{covder}
\hspace{-1.5em}
\nabla_{\widetilde C} \widetilde X^{\widetilde A_1 \widetilde A_2 \ldots \widetilde A_p}_{\widetilde B_1 \widetilde B_2 \ldots \widetilde B_q} \ =\ \Omega^{-(d_X-1)} (K^{\widetilde A_1}_{\ A_1}K^{\widetilde A_2}_{\ A_2} \ldots K^{\widetilde A_p}_{\ A_p})  
\, (\nabla_C X^{A_1 A_2 \ldots A_p}_{  B_1  B_2 \ldots B_q }  )\,
K^C_{\ \widetilde C}  (K^{ B_1}_{\ \widetilde B_1}K^{ B_2} _{\ \widetilde B_2}\ldots K^{ B_q}_{\ \widetilde B_q}) \;.
\nonumber
\\
\vphantom{e}
\end{align}
where $\nabla_{\widetilde A} X$ represents the frame covariant derivative of $X$ with respect to the conformally transformed field in the new basis, $\widetilde \varphi^{\widetilde A}$. Focusing on preserving property \eqref{confdef} first, we may construct a conformally-covariant field derivative as follows:
\begin{align}
\label{confcovfieldder}
X^{A_1 A_2\ldots A_p}_{B_1 B_2\ldots B_q;C} \ \equiv\ X^{A_1 A_2\ldots A_p}_{B_1 B_2\ldots B_q ,C} - \frac{w_X}{2} \frac{f_{,C}}{f} X^{A_1 A_2\ldots A_p}_{B_1 B_2\ldots B_q}\; .
\end{align}
Using this derivative, we may write down a Christoffel-like connection using $G_{AB}$ as a metric analogue~\cite{Sasaki:1995aw},
 \begin{align}
 \label{conn}
\Gamma^A_{BC}\ =\ \frac{G^{AD}}{2}\, \big(G_{DB;C}+ G_{CD;B} -G_{BC;D}\big)\;  .
 \end{align}
This construction ensures that $\Gamma^A_{BC}$ is conformally invariant, with $w_\Gamma = 0$. As a consequence, the conformally-covariant derivative can be extended so as to incorporate field reparametrizations, leading to a fully \emph{frame-covariant} field derivative defined as
\begin{equation}
\begin{aligned}
\label{fieldcovderdef}
\nabla_C X^{A_1 A_2\ldots A_p}_{B_1 B_2\ldots  B_q }\ \equiv\  X^{A_1 A_2\ldots A_p}_{B_1 B_2\ldots B_q;C} &+ \Gamma^{A_1}_{CD} X^{D A_2\ldots A_p}_{B_1 B_2\ldots B_q}   
+ \cdots \, +   \Gamma^{A_p}_{CD} X^{A_1 A_2\ldots D}_{B_1 B_2\ldots B_q}
\\
& - \Gamma^{D}_{B_1 C} X^{A_1 A_2\ldots A_p}_{D B_2\ldots B_q} 
- \cdots
- \Gamma^{D}_{B_q C} X^{A_1 A_2\ldots A_p}_{B_1 B_2 \ldots B_q} \;.
\end{aligned}
\end{equation}
It is then straightforward to check that this definition of the frame-covariant derivative satisfies the covariance condition specified in \eqref{covder}. Given~\eqref{fieldcovderdef}, it is possible to define a frame-covariant derivative ${\cal D}_\lambda$ with respect to any parameter $\lambda$ as follows:
\begin{align}
\label{paramcovder}
\mathcal{D}_\lambda X^{A_1 A_2 \ldots A_p}_{B_1 B_2\ldots B_q}\ 
&\equiv\  \frac{d \varphi^C}{d \lambda} \,
\nabla_C X^{A_1 A_2 \ldots A_p}_{B_1 B_2\ldots B_q}\; .
\end{align}

We now turn our attention to cosmology. With the aid of~\eqref{paramcovder}, we may extend the definition of the usual Hubble parameter $H$ to the \emph{covariant Hubble parameter}~$\mathcal{ H}$ by promoting the ordinary time derivative to a covariant derivative:
\begin{align}
\label{covhrate}
\mathcal{ H}\   \equiv\  \frac{\mathcal{D}_t  a}{a} \;,
\end{align}
where $dt = N_L d\tau$ is the {\em physical} cosmic time element.
Applying \eqref{paramcovder} to a quantity $X$ with no field indices, e.g.~$X = a$, 
the frame-covariant derivative $\mathcal{D}_\lambda X$ takes on the explicit form
\begin{align}
\label{dldef}
\mathcal{D}_\lambda X  \ &=\ X_{,\lambda} - \frac{w_X}{2} \frac{f_{,\lambda}}{f} X 
\nonumber\\
&=\ X_{,\lambda} +\ d_X {\dot \lambda}^{-1} H_L X 
\; .
\end{align}
In the above, we have eliminated the dependence on $f$ by making use of the fact that
\begin{align}
\label{lapsdof}
H_L\ =\ \frac{ \dot N_{L }}{N_L }\ =\ -\,\frac{\dot f }{2f}  \ .
\end{align}
The latter derives from the observation that $N_L^2 f$ is a constant by construction in the Einstein frame, and as such, invariant in all frames since it possesses scaling dimension 0. With the help of~\eqref{dldef} and~\eqref{lapsdof},
we find that~$\mathcal{H} = (H - H_L)/N_L$. In particular, we find that for $X = f$, \eqref{dldef} with $w_f = 2$ implies
\begin{align}
\label{ddf}
\mathcal{D}_\lambda f = 0\;
\end{align}
for any parameter $\lambda$. Equation~\eqref{ddf} leads to the statement of the indistinguishability between frames of a theory, which amounts to saying that no experiment or observation can discriminate between different values of the effective Planck mass squared $f$, at least at the classical level~\cite{Dicke:1961gz}.

As an example of \eqref{dldef}, it is illustrative to consider the second covariant time derivative of the fields $\varphi^A$. Setting $X^A \equiv \mathcal{D}_t \varphi^A$ in \eqref{paramcovder}, where
 \begin{align}
 \label{dtdef}
\mathcal{D}_t  \varphi^A  &= N_L^{-1} \big(\dot \varphi^A +  H_L \varphi^A\big) \; ,
\end{align}
we may calculate 
\begin{align}
 \label{dtdefsec}
\mathcal{D}_t  \mathcal{D}_t  \varphi^A   &=   N_L^{-1} \big( \dot X^A + \Gamma^A_{BC} X^B X^C \big) \;.
\end{align}
Employing~\eqref{paramcovder}, along with the frame-covariant quantities $G_{AB}$, $U$, and $\mathcal{H}$, it is now possible to rewrite the equations of motion \eqref{infeqFRW}--\eqref{acceleq} in a fully frame-covariant manner. In detail, \eqref{infeqFRW} becomes
 \begin{align}
 \label{invarinfleq}
&  \mathcal{D}_t \mathcal{D}_t{\varphi}^A
+    3\mathcal{H}     (\mathcal{D}_t {  \varphi}^A )
+   f   G^{AB} U_{,B}\ =\ 0\;. 
 \end{align}
Correspondingly, \eqref{friedeq} and \eqref{acceleq}  become
\begin{align}
\label{friedeqinv}
\mathcal{H}^2\ &=\   \frac{1}{3 } \left( \frac{G_{AB} (\mathcal{D}_t \varphi^A )(\mathcal{D}_t \varphi^B)}{2 } +  f U \right)\; , 
\\
\label{acceleqinv}
\mathcal{D}_t {\mathcal{H}} &= -  \frac{G_{AB} (\mathcal{D}_t \varphi^A )(\mathcal{D}_t \varphi^B)}{2}\ .
\end{align}
It is easy to verify that \eqref{invarinfleq}--\eqref{acceleqinv} reduce to their well-known forms for single-field inflation. We observe that \eqref{invarinfleq} resembles a geodesic equation with two external forces: (i) a drag term proportional to $\mathcal{D}_t \varphi^A$ and (ii) a conservative external force proportional to $U_{,A}$. This analogy to differential geometry will be further explored in Section~\ref{quantpert}. Most importantly for now, these equations are fully frame-covariant; each term transforms with exactly the same weight and Jacobian, as can be seen in Table \ref{tab:confweight}. By comparing \eqref{friedeqinv} and \eqref{acceleqinv} with the usual (minimal) form of the Friedmann and acceleration equations written in terms of the energy density $\rho$ and pressure $P$, we may also define the covariant comoving energy density~$\varrho$ and pressure $\mathcal{P}$ as
\begin{equation}
\begin{aligned}
\varrho \  &\equiv\   
 \frac{G_{AB}}{2} (\mathcal{D}_t \varphi^A) (\mathcal{D}_t  \varphi^B)\: +\:   f U\;,
  \\
\mathcal{P}\ 
&\equiv \  \frac{G_{AB}}{2}(\mathcal{D}_t \varphi^A) (\mathcal{D}_t  \varphi^B)\:  -\:   f U\;.
\end{aligned}
\end{equation}

 \begin{table}
\centering

\begin{tabular}{ lcc   }
$X$	 			& conformal weight ($w_X$)& scaling dimension ($d_X$)			 \\	
\hline
$dx^\mu$			&   $\hphantom{-}0$ 	&  $\hphantom{-}0$		 		\\
$d\varphi^A$		&   $\hphantom{-}0$ 	&  $\hphantom{-}1$				\\
$d\varphi_A$		&   $\hphantom{-}0$ 	&  $-1$						\\
$g_{\mu\nu}$		&   $-2$ 			&  $-2$	 					\\
$g^{\mu\nu}$		&   $\hphantom{-}2$ 	&  $\hphantom{-}2$				\\
$N_L,a$			&   $-1$ 			&  $-1$	 					\\
$\mathcal{H}$		&   $\hphantom{-}1$	&  $\hphantom{-}1$		 		\\
$f$				&   $\hphantom{-}2$    	&  $\hphantom{-}2$		 		\\ 
$G_{AB}$  			&   $\hphantom{-}0$	&  $-2$		 				\\
$G^{AB}$  			&   $\hphantom{-}0$	&  $\hphantom{-}2$				\\
$U$  				&  $\hphantom{-} 0	$	&  $\hphantom{-}0$		 		\\
$X^{A_1 A_2 \ldots A_p}_{B_1 B_2 \ldots B_q}$ 			& $w_X $			&  $w_X +p-q $	\\
$\nabla_AX^{A_1 A_2 \ldots A_p}_{B_1 B_2 \ldots B_q}$ 			& $w_X $			&  $w_X-1+p-q $	\\
$\mathcal{D}_\lambda X^{A_1 A_2 \ldots A_p}_{B_1 B_2 \ldots B_q}$	&$w_X-d_{\delta \lambda}$		&$w_X-d_{\delta \lambda} +p-q$ \\
\end{tabular}

 \caption{Conformal weights and scaling dimensions of various frame-covariant quantities.}
  \label{tab:confweight}
\end{table}

In this section, we have further developed the concept of frame covariance and recast the classical equations of motion which govern the dynamics of inflation into frame-covariant forms as given in \eqref{invarinfleq}--\eqref{acceleqinv}. Our aim in the next section is to examine how these equations may be used in order to determine the evolution of the perturbations which form the seeds for the observable cosmological anisotropies. 

\section{Frame Covariance in Multifield Inflation}
\label{quantpert}

The generation of anisotropies on the surface of last scattering is fundamentally a quantum phenomenon, as the cosmological perturbations that eventually source the profile of the Cosmic Microwave Background~(CMB)  are seeded by the correlation functions of the primordial perturbations of the metric and the scalar fields. These perturbations freeze as they cross the cosmological horizon in the single field scenario, thus leaving their imprint on the CMB~\cite{Weinberg:2003sw,Weinberg:2008nf}. In the presence of multiple fields, inflationary trajectories can be described as living on a manifold, which naturally leads to the decomposition of the perturbations into \emph{curvature} modes parallel to the inflationary trajectory and the perpendicular \emph{isocurvature} (or \emph{entropic}) modes. The picture is further complicated by the fact that the isocurvature modes, unlike the curvature modes, are not conserved after they exit the horizon \cite{Polarski:1994rz}. 

In this section, using well-established results from differential geometry, we introduce the concept of a \emph{field space} first formally identified by~\cite{Gong:2011uw}, and extend it to be fully covariant following our discussion in Section \ref{sec:frametrans}. We discuss the phenomenological impact of the entropy transfer between curvature and isocurvature modes before specializing to two-field inflation. Finally, we analyze the effects of selecting boundary conditions on the observed quantities.

\subsection{Differential Geometric Approach to Perturbations}
\label{diffgeom}
We begin by perturbing the metric and the scalar fields around their classical values. We can parametrize the first order perturbations of the metric in the so-called scalar-vector-tensor decomposition:
\begin{equation}
\label{frwdec}
\begin{aligned}
g_{\mu\nu} \, dx^\mu dx^\nu \ = &\hphantom{+}   \ N_L^2 \  (1+2\phi) \ d\tau^2  
\\
&+  \ 2 \ ( a N_L) \ ( \mathcal{D}_i B + B_i)\, d\tau\, dx^i
\\ 
&-  \ a^2 \Big[\big(1  -  2\psi \big)\delta_{ij}\  + \ \mathcal{D}_i \mathcal{D}_j E \ +\  \big(\mathcal{D}_i E_j - \mathcal{D}_j E_i\big) + h_{ij} \Big] \, dx^i \, dx^j.
\end{aligned}
\end{equation}
In this decomposition, $\phi$, $\psi$, $A$ and $B$ are the four scalar perturbations, $A_i$ and $B_i$ contain the four vector perturbations (with the constraint $\partial_i A^i = \partial_i B^i = 0$), and $h_{ij}$ contains the two tensor perturbations (with the constraint $h^j_j = \partial_i h^i_j = 0$).  The \emph{Bardeen potentials}~\cite{Bardeen:1980kt} are denoted by~$\Phi$ and $\Psi$. We may thus write in the Newtonian gauge:
\begin{align}
\label{metricpar}
g_{\mu\nu} dx^\mu dx^\nu\ =\ (1+2\Psi)N_L^2 \, dt^2\: -\: a^2 \big[ (1-2\Phi)\delta_{ij} + h_{ij}\big] dx^i dx^j,
\end{align}
where we focus only on scalar and tensor perturbations. The Bardeen potentials~$\Psi$ and~$\Phi$ are equal in the absence of anisotropic stress, which is the case when we are dealing with scalar fields. We may further define the frame-covariant extensions of the gauge-invariant \emph{Mukhanov-Sasaki} variables~\cite{Sasaki:1986hm,Mukhanov:1988jd} as follows:
\begin{align}
\label{sasakivar}
Q^A\ \equiv\ \delta \varphi^A + \frac{\mathcal{D}_t \varphi^A}{\mathcal{H}} \Phi\; .
\end{align}
The difference between the standard decomposition found in the literature \cite{White:2012ya, White:2013ufa} and the decomposition \eqref{frwdec}--\eqref{sasakivar} is the use of frame-covariant derivatives, which ensure that the metric perturbations are also frame-invariant.  Note that the frame invariance of $Q^A$ stems from the definition of the frame-covariant time derivative, the invariant Bardeen potentials and the covariant Hubble parameter $\mathcal{H}$, with $w_Q = w_\Phi = 0$ and $w_{\delta \varphi} = 0$. It is important to note that $\delta \varphi^A$ is independent of $\varphi^A$, which explains the difference in their conformal weight. Hence, if we were to decompose a scalar field into its background component~$\varphi^A$ and its variation~$\delta \varphi^A$, we would write $\varphi^A + M_P \delta\varphi^A$, ensuring dimensional consistency according to~Table~\ref{tab:confweight}.

The concept of the \emph{field space metric} arises naturally when considering the inflationary trajectories satisfying the scalar field equation~\eqref{invarinfleq}, as the fields $\varphi^A$ take on the roles of coordinates~\cite{Gong:2011uw, Elliston:2012ab}. The simplest choice for the metric that generates~\eqref{invarinfleq} in the absence of external forces is $G_{AB}$, which is defined in~\eqref{eq:GAB}, which also naturally leads to the definition of the field-space connection $\Gamma^{A}_{BC}$ as given in~\eqref{conn}. The metric $G_{AB}$ is assumed to have an inverse $G^{AB}$, and both can be used to raise and lower indices of vectors and covectors living in the manifold.  The utility of the field space formalism becomes apparent once isocurvature perturbations enter the picture, since it is preferable to decompose perturbations in their curvature and isocurvature components living in the tangent and cotangent spaces respectively. In order to rewrite $Q^A$ in terms of curvature and isocurvature perturbations, we use a set of \emph{frame fields} $e_a^A$ by the following property: 
\begin{align}
\label{frameprop}
G_{AB}\ \equiv\ e^\alpha_A e^\beta_B \delta_{\alpha\beta}\;,
\qquad
G^{AB}\ \equiv\ e_\alpha^A e_\beta^B \delta^{\alpha\beta}\;.
\end{align}
Using these frame fields, we may switch from the global field space basis denoted by $A, B, \ldots$ to the local basis:
\begin{align}
\label{decomp}
Q^\alpha\ =\ Q^A e_A^\alpha\;,
\hspace{3.2em}
Q^A\ =\ Q^\alpha e_\alpha^A\;.
\end{align}
Local basis indices $\alpha, \beta, \ldots$ run over $\{ \sigma,s_1,s_2,\ldots \}$, where $\sigma$ corresponds to the curvature component, and $i \in  \{s_1, s_2, \ldots\}$, corresponding to the isocurvature components. 
This decomposition allows us to write the inflaton equations of motion \eqref{invarinfleq} as
\begin{align}
\label{infleqsigma}
\mathcal{D}_t \mathcal{D}_t \sigma
+    3\mathcal{H}     (\mathcal{D}_t \sigma)
+   f    U^{,\sigma}\ =\ 0 \; .
\end{align}
We may explicitly write
\begin{align}
\label{framefields}
e_\sigma^A\  &=\ \frac{\mathcal{D}_t \varphi^A}{ \mathcal{D}_t \sigma}\ ,
&
e_{s_1}^A \ &=\ - \frac{(s_1)^{AB} \, U_{,B}}{ \sqrt{(s_1)^{AB} \, U_{,A} U_{,B} }}\ =\ \frac{\omega^A}{\omega}\ ,
\end{align} 
where we focus on the so-called ``first'' isocurvature mode $Q^{s_1}$, which is defined to be parallel to the component of acceleration perpendicular to the tangent space and
\begin{align}
 {(s_1)}^{A B}\ \equiv\ G^{AB} - e_{\sigma}^A  e_{\sigma}^B\;,
\end{align}
We have defined the frame field  $e_{s_1}^A$ associated with the first entropic mode, where the \emph{turn rate} $\omega$ is the field space magnitude of the \emph{acceleration vector} $\omega^A$ which effectively measures the rate of change between geodesics in curved space and is given by
\begin{align}
\label{turnrate}
\omega^A\ =\  \mathcal{D}_N \left(\frac{\mathcal{D}_t \varphi^A}{\mathcal{D}_t \sigma} \right).
\end{align}
We single out the first isocurvature mode because it directly couples to $\mathcal{R}$, whereas the remaining isocurvature modes do not~\cite{McAllister:2012am}, meaning that all entropy transfer between curvature and isocurvature modes can be traced back to the coupling between $Q^\sigma$ and $Q^{s_1}$. We may intuitively understand this by visualizing the trajectory as it turns in field space: since the first isocurvature mode is by definition the only component parallel to the acceleration vector, it is the only mode which can source the curvature mode. It is possible to further define a projection operator perpendicular to both the acceleration vector $\omega^A$ and the tangent vector $e^A_\sigma$ through ${(s_2)}^{A B}\ \equiv\ (s_1)^{A B} - e_{s_1}^A  e_{s_1}^B$. This may be used to define the second entropic mode, which will couple only to the first entropic mode as the trajectory turns in the submanifold spanned by~${(s_2)}^{A B}$. Repeating this procedure, we have a hierarchy of modes, all of which couple to their immediate neighbours: the curvature mode couples to the first isocurvature mode, which couples to the second isocurvature mode, and so forth.

In order to make contact with phenomenology, we now define the observationally relevant quantities of interest to us, which are the \emph{comoving curvature perturbation} $\mathcal{R}$ and the \emph{comoving isocurvature perturbations} $\mathcal{S}^{(i)}$:
\begin{align}
\label{curvpert}
\mathcal{R}\ \equiv\ \frac{\mathcal{H}}{\mathcal{D}_t \sigma} Q^\sigma, \qquad
\mathcal{S}^{(i)}\ \equiv\ \frac{\mathcal{H}}{\mathcal{D}_t \sigma} Q^{s_i}.
\end{align}
Both $\mathcal{R}$ and $\mathcal{S}^{(i)}$ are gauge- and frame-invariant. The curvature perturbation $\mathcal{R}$ is of particular interest to us, as it remains constant on superhorizon scales and sources the (dimensionless) observable scalar spectrum $P_\mathcal{R}$ through its two-point function:
 \begin{align}
 \label{scalpowspec}
\frac{2\pi^2}{p^3} P_\mathcal{R} \, \delta^{(3)}({\bf p} + {\bf q})\ \equiv\ \braket{\mathcal{R}_{\bf p}|\mathcal{R}_{\bf q}} \,.
\end{align}
We may find an expression for $P_\mathcal{R}$ in terms of the background by solving and canonically quantizing the perturbed equations of motion \cite{Sasaki:1995aw}:
 \begin{align}
 \label{perteq}
\mathcal{D}_t \mathcal{D}_t Q_{\bf p}^A + 3 \mathcal{H}\, ( \mathcal{D}_t  Q_{\bf p}^A) + \frac{p^2}{a^2} Q_{\bf p}^A +  M^A_{\ B} Q_{\bf p}^B \ - \frac{1}{N_L a^3} \mathcal{D}_t \left[\frac{N_L a^3}{\mathcal{H}} (\mathcal{D}_t \varphi_A)  (\mathcal{D}_t \varphi_B) \right]=\ 0\;,
\end{align}
where the frame covariant mass matrix $M_{AB}$ is given by
 \begin{align}
 \label{massmatrix}
M_{AB}\ \equiv\  f (\nabla_A \nabla_B U) + R_{AMBN} (\mathcal{D}_t \varphi^M) (\mathcal{D}_t \varphi^N) \, 
\end{align}
and $R_{AMNB}$ is the field-space covariant Riemann tensor defined through the Christoffel-like connection $\Gamma^{A}_{BC}$ given in~\eqref{conn} in the usual way. After imposing the usual commutator relations on the ladder operators and using the flat Bunch-Davies vacuum condition for very early times, we arrive at the following expressions for the two-point function of the (dimensionless) scalar power spectrum at horizon crossing~\cite{ Stewart:1993bc,Nakamura:1996da} and tensor power spectrum~$P_T$:
 \begin{align}
 \label{scal}
 P_{\mathcal{R} } \ =\ \frac{ \mathcal{H}^2}{8\pi^2 f(\varphi) \bar \epsilon_H}\;,
 \qquad
 P_T\ =\ \frac{2  }{ \pi^2}  \frac{\mathcal{H}^2}{f(\varphi)}\ .
\end{align}
where we define the \emph{frame-invariant Hubble slow-roll parameters} $\bar \epsilon_H$ and $\bar \eta_H$ as
  \begin{align}
  \label{srp}
\bar \epsilon_H \ \equiv\   -\frac{\mathcal{D}_t \mathcal{H}}{\mathcal{H}^2}\; ,
\qquad 
\bar \eta_H\  \equiv\  \, \frac{\mathcal{D}_t \bar \epsilon_H}{\mathcal{H} \,\bar \epsilon_H}\;.
 \end{align}
The expressions for the power spectra  \eqref{scal} are derived under the \emph{slow-roll approximation}, which we may write in a covariant form as
\begin{align}
\label{sra}
\mathcal{D}_t \mathcal{D}_t {\varphi^A}  \ll \mathcal{H} (\mathcal{D}_t  \varphi^A )\;.
 \end{align}
 Any field $\varphi^A$ that satisfies this hierarchy contributes to inflation. Fields that happen not to obey~\eqref{sra} effectively act as spectator fields, such as those that appear in the curvaton scenario~\cite{Lyth:2001nq}. It is worth noting that this procedure is identical to the usual treatment in the literature, the only difference being the use of frame-covariant derivatives which ensures that the resulting expressions for the dimensionless spectra are frame-invariant.

\subsection{Frame-Covariant Observable Quantities}

It is easy to see that \eqref{srp} reduces to its standard form in the Einstein frame where~$\mathcal{H} \rightarrow H$, $f\rightarrow M_P^2$ and $\bar\epsilon_H \rightarrow \epsilon_H$. However, in the multifield case, isocurvature modes evolve outside the horizon and because they are coupled to the curvature modes, the power spectrum \eqref{scal} evaluated at horizon exit differs from the observable spectrum at horizon re-entry \cite{Wands:2002bn}. 
As noted in Subsection \eqref{diffgeom}, we expect the perturbations to form a hierarchy in which every mode couples only to its immediate neighbours. Indeed, the equations of motion for the perturbations in terms of the curvature perturbations $\mathcal{R}$ and~$\mathcal{S}^{(i)}$ in the long wavelength limit, where the coupling between~$\mathcal{R}$ and~$\mathcal{D}_t \mathcal{R}$ vanishes~\cite{Kaiser:2012ak}, we find
 \begin{equation}
 \begin{aligned}
 \label{superhorizon}
\mathcal{D}_t \mathcal{R} &= A  \mathcal{H} \mathcal{S}^{(1)},
\\
\mathcal{D}_t \mathcal{S}^{(1)} &= B_1 \mathcal{H} \mathcal{S}^{(2)},
\\
&\hspace{0.55em}\vdots
\\
\mathcal{D}_t \mathcal{S}^{(n-2)} &= B_{n-2}\mathcal{H} \mathcal{S} ^{(n-1)},
\\
\mathcal{D}_t \mathcal{S}^{(n-1)} &= B_{n-1}\mathcal{H} \mathcal{S} ^{(n-1)},
\end{aligned}
\end{equation}
where $A(t)$ and $B_1(t), B_2 (t), \ldots, B_{n-1}(t)$ are model-dependent parameters.
Since they are field space scalars, the above system of equations \eqref{superhorizon} is fully frame-invariant. Solving the system of equations \eqref{superhorizon}, we find that for times $t >t_*$, where $t_*$ is the time of horizon exit for a pivot scale of cosmological interest with wavenumber $k_* \approx 0.002 \text{ Mpc}^{-1}$, the perturbations evolve according to
\begin{equation}
\begin{aligned}
\begin{pmatrix}
\mathcal{R}(t)
\\
\mathcal{S}^1(t)
\\
\mathcal{S}^2(t)
\\
\vdots
\\
\mathcal{S}^{n-2}(t)
\\
\mathcal{S}^{n-1}(t)
\end{pmatrix}
=
\begin{pmatrix}
1 & T_\mathcal{RS} & 0 & 0 & \cdots   & 0
\\
0 & 0 & T_{12}& 0 & \cdots  & 0
\\
0 & 0 & 0   & T_{ 23} & \cdots  & 0 
\\
\vdots & \vdots&\vdots & \vdots&  \ddots & \vdots  
\\ 
0 & 0& 0& 0& \cdots    & T_{ (n-2) (n-1) }
\\
0 & 0& 0& 0  & \cdots  & T_{ (n-1) (n-1) }
\end{pmatrix}
\begin{pmatrix}
\mathcal{R}_*
\\
\mathcal{S}^{(1)}_*
\\
\mathcal{S}^{(2)}_*
\\
\vdots
\\
\mathcal{S}^{(n-2)}_*
\\
\mathcal{S}^{(n-1)}_*
\end{pmatrix},
\end{aligned}
\end{equation}
where $\mathcal{R}_* = \mathcal{R}(t_*)$, $\mathcal{S}^{(i)}_* = \mathcal{S}^{(i)}(t_*)$, and $T_{ij}=T_{ij}(t_*,t)$ denotes the transfer function between the $i$th and $j$th isocurvature modes. The formal solutions to the superhorizon equations~\eqref{superhorizon} are 
\begin{equation}
\label{transfunct}
 \begin{aligned}
 T_\mathcal{RS}(t_*, t) &=  \int_{t^*}^t dt'  \; T_{12} (t_*, t') A(t') \mathcal{H} ,
 \\
 T_{12} (t_*, t) &= \exp \left[ \int_{t^*}^t dt'  \;  T_{23 } (t_*, t')   B_1(t')  \mathcal{H}\right],
\\
&\hspace{0.55em}\vdots
\\
T_{(n-2)(n-1)} (t_*, t) &= \exp \left[ \int_{t^*}^t dt'  \; B_{n-2}(t')  \mathcal{H}\right],
\\
T_{(n-1)(n-1)} (t_*, t) &= \exp \left[ \int_{t^*}^t dt'  \; B_{n-1}(t')  \mathcal{H}\right].
\end{aligned}
\end{equation}
 To first order, we may neglect the corrections to the power spectrum that are of the order of the slow-roll parameters \cite{Byrnes:2006fr} and hence write
\begin{align}
\label{PR1st}
P_{\mathcal{R}}(t)\ &=\ \Big[1  +  T^2_\mathcal{RS}(t_*,t)\Big] P_{\mathcal{R}  }(t_*)\;
 =\ P_{\mathcal{R}}(t)  \cos^{-2} \Theta\;,
\end{align}
where the effect of the coupling between curvature and isocurvature modes may be absorbed into the \emph{transfer angle} $\Theta$.  We may also define the isocurvature fraction $\beta_{\rm iso}$ \cite{Kaiser:2013sna, Schutz:2013fua}:
\begin{align}
\beta_{\rm iso} \equiv  \frac{  \sum_{i=1}^{n-1}   P_{\mathcal{S}^{(i)} } }{P_\mathcal{R}+ \sum_{i=1}^{n-1}   P_{\mathcal{S}^{(i)} }} = \frac{T_{12}^2 \; + \; T_{23}^2 \; + \ldots + \; T_{(n-1)(n-1)}^2}{1 \; + T_\mathcal{RS} + \; T_{12}^2 \; + \; T_{23}^2 \; + \ldots + \; T^2_{(n-1)(n-1)}},
\end{align}
where $ P_{\mathcal{S}^{(i)}} $ is defined through a relation similar to \eqref{scalpowspec}.

We have studied the generic features of the entropy transfer between curvature and isocurvature modes. We will apply our analysis to a generic two-field scenario in the next section and provide numerical results for concrete models in Section \ref{specmod}. As for now, we may  write down concise and fully frame-invariant expressions for a number of relevant inflationary observables. The power spectra $P_\mathcal{R}$ and $P_T$ can be used to define the following standard inflationary observables. These are the \emph{scalar spectral index} $n_\mathcal{R} $, the \emph{tensor spectral index}~$n_T $, and \emph{tensor-to-scalar ratio} $r$:
 \begin{align}
 \label{odef1}
n_\mathcal{R} - 1\equiv\left.\frac{d \ln  P_\mathcal{R} }{d\ln k}\right\vert_{k =  a \mathcal{H} } ,
\qquad
n_T \equiv\left.\frac{d \ln  P_T }{d\ln k}\right\vert_{k =  a \mathcal{H}},
\qquad
r \equiv     \frac{  P_T }{  P_\mathcal{R} },
\end{align}
where we evaluate every parameter at the time of horizon exit $k=a \mathcal{H}$ in all expressions for the observables.
We also define the \emph{runnings} of the spectral indices as follows:
\begin{align}
\label{odef2}
\alpha_\mathcal{R} &\equiv \left.\frac{d    n_\mathcal{R} }{d\ln k}\right\vert_{k =    a \mathcal{H}}\; ,
\qquad
\alpha_T   \equiv \left.\frac{d    n_T }{d\ln k}\right\vert_{k =    a \mathcal{H}}\;,
\end{align}
as well as the \emph{non-linearity parameter}~$f_{NL}$, which is defined is defined through the three point correlation function for $\mathcal{R}$~\cite{Komatsu:2001rj,Huang:2008bg,Ganc:2011dy} and is given by~\cite{Elliston:2012ab,Byrnes:2012sc}:
 \begin{align}
 \label{fNL}
f_{NL}\ =\ \frac{5}{6} \frac{  N^{,A} N^{,B} (\nabla_A \nabla_B N)}{(  N_{,A} N^{,A})^2}\;,
\end{align}
where the number of \emph{e-folds} $N$ is related to time via $dN = - \mathcal{H} \, dt$ and the frame-covariant derivative $\mathcal{D}_N$ with respect to the number~$N$ of $e$-folds is defined with the help of \eqref{paramcovder}. 
Note that in the minimal case where the field space is not curved, the covariant field derivatives in~\eqref{fNL} would be replaced by ordinary derivatives~\cite{Lyth:2005fi}. The expressions for these observables to first slow-roll order can be written by applying the definitions \eqref{odef1} and \eqref{odef2} to \eqref{scal}:
\begin{align} 
\label{observables}
n_\mathcal{R}\  &=\  1 - 2 \bar \epsilon_H -  \bar \eta_H - \mathcal{D}_N \ln \big(1+ T_\mathcal{RS}^2 \big),
&
n_T\  &=\  -2 \bar \epsilon_H\;,
&
r\  &=\  16 \bar  \epsilon_H \cos^{2}\Theta\; ,
\nonumber\\
\alpha_\mathcal{R}\   &=\     
 - 2 {\bar \epsilon}_H  {\bar\eta}_H  - {\bar \eta}_H {\bar \xi}_H
 + \mathcal{D}_N\mathcal{D}_N \ln \big(1 + T_\mathcal{RS}^2\big)\;,
 &
\alpha_T\  &=\ 
-2\bar\epsilon_H\bar\eta_H\;.
\end{align}
We have defined $\bar \xi_H$ as part of a hierarchy of Hubble parameters, whose terms appear when we calculate higher-order runnings of the inflationary observables. In detail, we have
\begin{equation}
\label{hsrphierarchy}
\begin{aligned}
\bar \epsilon_{H,1} \ \equiv \ \bar \epsilon_H\;,
\quad
\ldots\,,
\quad
\bar \epsilon_{H,n}\   \equiv \ - \mathcal{D}_N \ln \epsilon_{H,n-1}\;,
 \end{aligned}
 \end{equation}
with $\bar \epsilon_{H,2} = \bar \eta_H $ and $\bar\epsilon_{H,3} = \bar \xi_H$. The expressions \eqref{observables} are exactly similar to the ones reported in the literature, the only difference being the use of the manifestly frame-invariant slow-roll parameters (which reduce to the standard parameters in the Einstein frame).  

Employing the relations \eqref{hsrphierarchy} for the inflationary parameters, we find a relation that holds for all multifield scalar-curvature models of inflation:
 \begin{align}
\label{2ndconseq}
\Big[1 + n_T - n_\mathcal{R} -  \mathcal{D}_N\mathcal{D}_N \ln \big(1 + T_\mathcal{RS}^2\big) \Big] r \ =\  -8\alpha_T \cos^2\Theta\; .
\end{align}
This is a consistency relation which can become a significant observational test for {\em both} single and multifield inflation, if a large fraction~$r$ of tensor perturbations is  detected in future. We note that equivalent expressions have been reported in \cite{vanTent:2003mn} and \cite{Byrnes:2006fr}.

In the slow-roll approximation, the deviation from the geodesic in field space is small. The latter results from setting $\mathcal{D}_t \mathcal{D}_t \varphi^A = 0$ in~\eqref{invarinfleq} to zero, which implies
  \begin{align}
  \label{src}
  3 \mathcal{H}   (D_t \varphi^A) +   f   U^{,A} &= 0\; .
\end{align}
This equation for $\varphi^A$ is known as the \emph{inflationary attractor}, which essentially defines a class of trajectories that the scalar fields will approach towards regardless of initial conditions. We are now in the position to define the frame-invariant \emph{potential} slow-roll parameters, whose defining feature is that they reduce to the Hubble slow-roll parameters in the slow-roll approximation as $\bar \epsilon_{U,n} \approx \bar \epsilon_{H,n}$.  Using \eqref{src}, we define
  \begin{align}\label{psrp}
\bar \epsilon_U &\equiv   \frac{1 }{2 }  \frac{G^{AB}   U _{,A}  U_{,B}}{ U^2},
&
\bar \eta_U &\equiv  -\frac{(\bar \epsilon_U)_{,A}}{\bar \epsilon_U}  G^{AB} \frac{U_{,B}}{U},
&
\bar \xi_U &\equiv  -\frac{(\bar \eta_U)_{,A}}{\bar \eta_U}  G^{AB} \frac{U_{,B}}{U}.
\end{align}
These parameters are part of a potential slow-roll parameter hierarchy, in exact analogy with the Hubble slow-roll parameter hierarchy given in \eqref{hsrphierarchy}. More explicitly, we have
\begin{equation}
\label{xisrp}
\begin{aligned}
\bar\epsilon_{U,1} \ \equiv \ \bar\epsilon_U\;,
\quad
\ldots\,,
\quad
\bar \epsilon_{U,n} \  \equiv\ -\frac{(\bar \epsilon_{U,n-1})_{,A}}{\bar \epsilon_{U,n-1}}  G^{AB} \frac{U_{,B}}{U}\ ,
\end{aligned}
\end{equation}
where $\bar\epsilon_{U,2} \equiv \bar\eta_U$ and  $\bar\epsilon_{U,3} \equiv \bar\xi_U$. Given the definitions of the potential slow-roll parameters in \eqref{psrp}, we may easily write down concise expressions for the cosmological observables in terms of $\bar \epsilon_U$, $\bar \eta_U$, and $\bar \xi_U$:
\begin{equation}
\label{potobs}
\begin{aligned} 
n_\mathcal{R} &=  1 - 2 \bar \epsilon_U +  \bar \eta_U - \mathcal{D}_N \ln \big(1+ T_\mathcal{RS}^2\big)\;,
&
n_T &=  -2\bar\epsilon_U\;,
&
r &= 16 \bar \epsilon_U (\cos\Theta)^2\;,
\\
\alpha_\mathcal{R}  &=     
 - 2 {\bar \epsilon}_U  {\bar\eta}_U  - {\bar \eta}_U {\bar \xi}_U +  \mathcal{D}_N \mathcal{D}_N \ln \big(1+ T_\mathcal{RS}^2\big)\;,
&
\alpha_T &= 
-2\bar\epsilon_U\bar\eta_U\; .
\end{aligned}
\end{equation}
Likewise, the scalar and tensor power spectra read as follows:
 \begin{align}
{P_{\mathcal{R}}} \ =\ \frac{1}{24\pi^2} \frac{U}{\bar \epsilon_U} \, \cos^{-2} \Theta\; ,
\qquad
{P_{T}}\  =\ \frac{2 }{ 3\pi^2}\, U \; .
\end{align}
In the same vein, a simple expression for $f_{NL}$ may be obtained by substituting the following expression for $N_{,A}$ in \eqref{fNL}:
 \begin{align}
N_{,A}\ =\ \frac{U\, U_{,A}}{U_{,B}\, U^{,B}}\;.
\end{align}
We have thus defined two hierarchies of manifestly frame-invariant slow-roll parameters which we may use to write down expressions for the cosmological observables.

\subsection{Isocurvature Effects in Two-Field Models}
\label{subsec:twofield}

In this subsection, we will analyze the effects of entropy transfer between the curvature and isocurvature modes by specializing to two-field models.  Turning our attention to the entropy transfer between the modes, we find that solving the superhorizon equations of motion is in general very difficult. We may, however, derive approximately analytic results for two-field models. In this case, the isocurvature modes are fully encoded in $\mathcal{S} \equiv \mathcal{S}^{(1)}$, and the superhorizon equations of motion simplify to~\cite{Lalak:2007vi,vandeBruck:2016rfv}
\begin{equation}
\label{superhorizontwofield}
\begin{aligned}
\mathcal{D}_N \mathcal{R}\ &=\  -A(N)\, \mathcal{S}\;, \\
\mathcal{D}_N \mathcal{S}\ &=\ - B(N)\,   \mathcal{S}\; ,
\end{aligned}
\end{equation}
where $B_1 (t) = B(t)$. The solution to \eqref{superhorizontwofield} is given by means of the transfer functions~\eqref{transfunct}, which for two-field inflation take on the form
\begin{equation}
\label{transfeqnew}
\begin{aligned}
 T_{\mathcal{RS}}(N_*,N)\ &=\  -\int_{N_*}^N dN'  \,  A(N') \, T_{SS}(N_*,N')\; ,
 \\
 T_{\mathcal{SS}}(N_*,N)\ &=\  \exp\left[ - \int_{N_*}^N  dN' \, B(N')\right]\; .
\end{aligned}
\end{equation}
In two-field inflation, the parameters $A$ and $B$ are found to be \cite{Kaiser:2012ak}
\begin{equation}
\label{transparam}
\begin{aligned}
A(\varphi)\ &=\  2\omega\; ,
 \\
B(\varphi)\ &=\ -2 \bar \epsilon_H - \bar\eta_{ss} + \bar\eta_{\sigma\sigma}-\frac{4}{3} \omega^2\;,
\end{aligned}
\end{equation}
where the magnitude $\omega$ of the acceleration vector $\omega^A$ may be found through \eqref{turnrate} and the directional slow-roll parameters $\bar \eta_{AB}$ are defined with the help of the mass matrix given in~\eqref{massmatrix}, 
\begin{align}
\label{etadef}
  \bar \eta_{AB} \ \equiv\  \frac{M_{AB}}{fU}   \ \approx \   \frac{\nabla_A \nabla_B U}{U} + \frac{ 2 \bar \epsilon_U}{3 } R_{A\sigma B\sigma}\, .
\end{align}
We may switch between the global basis and the local basis through the frame fields given in \eqref{framefields}:
\begin{align}
  \label{eta1}
  \bar \eta_{\sigma\sigma} \ &=\ M_{AB}  e^A_\sigma e^B_\sigma \ \approx\   (\ln U)^{,A} (\ln U)^{,B} \frac{ \nabla_A \nabla_B  U   }{U}\ , 
  \\
  \label{eta2}
  \bar \eta_{ss}\ &=\  M_{AB}  e^A_s e^B_s\ \approx\ \frac{\omega^A \omega^B  }{\omega^2}  \frac{ \nabla_A \nabla_B  U   }{U}\ + \frac{\bar \epsilon_U}{3}  S   \ .
\end{align}
In arriving at the last equality in \eqref{eta1}, we made use of the fact that $e^A_\sigma$ given in \eqref{framefields} becomes $e^A_\sigma \approx (\ln U)^{,A}$ in the slow-roll approximation. 
Similarly, the slow-roll approximated expression on the RHS of~\eqref{eta2} is obtained by first noticing that $e^A_s = \omega^A/\omega$, and then that $S = 2R_{s \sigma s\sigma}$ is the Ricci scalar of the two-dimensional field space. In the  slow-roll regime, $\omega^A$ may be approximated to second order by rewriting the RHS of~\eqref{turnrate} as
\begin{align}
  \label{turnrateapprox}
 \omega^A\ &=\  (\ln U)^{,B}\ \nabla_B \left[\frac{(\ln U)^{,A}}{\sqrt{2 \bar\epsilon_U}}\right] .
\end{align}
In the same slow-roll regime, we may also compute the single and double covariant derivatives${}$ of the transfer function~$T_{\mathcal{RS}}$ with respect to $N$ by simply differentiating their integral forms~\eqref{transfeqnew} to find
\begin{equation}
\label{transder}
\begin{aligned}
\mathcal{D}_N T_{\mathcal{RS}} \ &=\  A_* + B_* T_\mathcal{RS}\;,
\\
\mathcal{D}_N \mathcal{D}_N  T_{\mathcal{RS}} \ &=\ A_* B_* + B_*^2 T_\mathcal{RS}\;.
\end{aligned}
\end{equation}
In this way, we can express all inflationary observables entirely in terms of the model functions\-
of the theory.

We may analytically perform the integrals in \eqref{transfeqnew} by assuming that the slow-roll parameters are slowly varying after horizon exit in the so-called \emph{constant slow-roll approximation}, such as in \cite{DiMarco:2005nq}. This is a crude approximation to the full solution, as the slow-roll parameters do evolve beyond the horizon. For most realistic inflation models, the turn rate $\omega$ peaks at~20 to 30 e-folds, which is well beyond the horizon exit at $N=60$. Therefore, this assumption will generally underestimate the amount of entropy transfer, providing a conservative value for $T_\mathcal{RS}$. Hence, under the assumption of constant slow-roll, we may now evaluate the parameters $A$ and $B$ at horizon crossing. From \eqref{transfeqnew}, we find
\begin{align}
\label{TSSapprox}
 T_{\mathcal{SS}} (N_*, N)\ &=\   e^{-B_*(N-N_*)}.
\end{align}
Substituting the last expression for $T_\mathcal{SS}$ in \eqref{transfeqnew}, we find
\begin{align}
\label{TRSold}
 T_{\mathcal{RS}}(N_*, N)\ &=\ \frac{A_* }{B_* } \Big[e^{ -B_*  (N-N_*) } - 1 \Big]\ .
\end{align}
We may further improve the constant slow-roll approximation by modelling the evolution\- of~$A_*$ as a linearly increasing function from zero to its maximum value $A_\text{max}$ in the interval: $N \in [0, N_\text{max}]$, and a linearly decreasing function from $A_\text{max}$ to $A_*$ in the interval: $N \in [ N_\text{max}, N_*]$, where $A(N_{\rm max}) = A_\text{max}$. Substituting this approximate form for $A$ in \eqref{transfeqnew} along with \eqref{TSSapprox}, we find that the transfer function at the end of inflation $N=0$ becomes 
 \begin{align}
 \label{TRSnew}
 T_{\mathcal{RS}}(N_*, 0)\ &=\ 
 \frac{\left(e^{\frac{B_* N_*}{2}}-1\right) \left\{2 A_\text{max} \left(e^{\frac{B_*N_*}{2}}-1\right)+ A_* \left[B_* N_*+e^{\frac{B_* N_*}{2}} (B_* N_*-2)+2\right]\right\}}{B_*^2 N_*}
 \ ,
\end{align}
where $N_\text{max} \approx N_*/2$ is a good approximation that will be used for most models of interest. In addition, we observe that in the limit~$A_\text{max}\rightarrow A_*$, the transfer function in~\eqref{TRSnew} reduces to one of~\eqref{TRSold}. We note that for scales whose horizon crossing $N = N_*$ coincides with the end of inflation~$N=0$, the transfer angle~$\Theta$ vanishes, as there is no time to generate entropy transfer. 

We conclude this subsection by commenting on the amplification and transfer of iso\-curvature perturbations in models with a curved field space. As seen in \eqref{transfeqnew}~and~\eqref{transparam}, entropic transfer occurs only if $\omega \ne 0$. We may relate the turn rate $\omega$ to the external forces
by rewriting $\omega^A$ in \eqref{turnrate} through the inflaton equations of motion \eqref{invarinfleq} and \eqref{infleqsigma}:
 \begin{align}
 \label{omegageodesic}
\omega^A\ &=\ -\,\frac{1}{\mathcal{H}} \frac{(\mathcal{D}_t\mathcal{D}_t \varphi^A) \mathcal{D}_t \sigma - \mathcal{D}_t\varphi^A (\mathcal{D}_t\mathcal{D}_t \sigma) }{(\mathcal{D}_t \sigma)^2}\ 
 =\ \frac{1}{\mathcal{H}} 
\frac{    e^A_{s_i} U^{,s_i} }{ \mathcal{D}_t \sigma}\ .
\end{align}
We thus observe that a non-zero turn rate occurs if and only if the force generated by the frame-invariant potential $U$ has a non-vanishing component perpendicular to the inflationary trajectory.  Such a force can only exist due to the potential, since the drag force $-3\mathcal{H}(\mathcal{D}_t \varphi^A)$ is always in the direction of the tangent vector. Hence, entropic transfer is not affected by a non-zero field-space Riemann curvature, $R_{A\sigma B\sigma}$, since neither the conservative force $-f U_{,A}$ nor the field-space geodesics are 
directly influenced by $R_{A\sigma B\sigma}$.

Unlike the entropic transfer, the amplification of isocurvature modes does depend on the field-space curvature $R_{A\sigma B\sigma}$. For two-field models, we observe that, even for a vanishing acceleration vector $\omega^A$, the field-space Ricci scalar, $S = 2R_{s \sigma s\sigma}$, will also contribute to the generation of isocurvature perturbations through $\bar \eta_{ss}$ [cf.~\eqref{eta2}]. For a small turn rate $\omega$, the generation of isocurvature modes will generically be amplified when the inflationary trajectory goes over a hilltop for which~$\bar \eta_{ss} < 0$, producing a tachyonic instability. Instead, the isocurvature modes will be suppressed when inflating in a valley with~$\bar \eta_{ss} > 1$. If~$0~<~\bar\eta_{ss}~<~1$, the generation of isocurvature modes will depend on the initial conditions~\cite{Kaiser:2013sna,Schutz:2013fua}. In general, we observe from~\eqref{eta2} that in the deep slow-roll regime, where~$\bar\epsilon_U,\ \bar\eta_U \ll 1$, the generation of entropic modes will be almost independent of $S$.  As we approach the end of inflation, $\bar\epsilon_U$ approaches unity and the effect of a negative field-space curvature $S$ can no longer be neglected, which may give rise to an enhanced entropy production for $\bar \eta_{ss}<0$.  Finally, we remark that both the generation of entropy and its transfer are controlled by the frame-covariant parameters $G_{AB}$ and $U$, and are in no way induced by conformal transformations to any particular~frame.
 
\subsection{Stability of Boundary Conditions}
\label{subsec:stability}

The last ingredient in our analysis pertains to the boundary conditions of the scalar fields. In this subsection, we discuss the dependence of the observable quantities on the choice of inflationary trajectory in field space, as well as the sensitivity of trajectories to the boundary conditions at the end of inflation. The end of inflation occurs when the comoving horizon~$(a \mathcal{H})^{-1}$ stops decreasing, which, in the Einstein frame, corresponds to the \emph{end-of-inflation} condition $  \epsilon_H \, = \, 1$, where $\epsilon_H \equiv - \dot H/H^2$. Slow-roll inflation carries the additional assumption that $\epsilon_H$ is slowly varying, which means that $\eta_H \equiv \dot \epsilon_H/(H \epsilon_H)$ must also be a small parameter. This is reflected in the \emph{slow-roll end-of-inflation condition} $\max(  \epsilon_H, \left\vert \eta_H \right\vert )   =  1 $.
These conditions may be easily made frame-invariant by using the invariant Hubble slow-roll parameters as follows:
\begin{align}
   \label{maxEF}
  \bar \epsilon_H \, = \, 1\,,
\qquad
\max(\bar \epsilon_H, \left\vert \bar \eta_H \right\vert)   \, =   \, 1\, .
\end{align}
Thus, the number of e-folds is frame-invariant since $t_0$ is also frame-invariant:
\begin{align}
\label{efolds}
N(t,t_0) = - \int_{t_0}^t dt' \, \mathcal{H}(t') \; .
\end{align}
We may express the conditions \eqref{maxEF} as constraints on the scalar fields through the potential slow-roll parameters by writing them as
\begin{align}
\label{maxgen}
\bar \epsilon_U  \, = \, 1\,,
\qquad 
\max(\bar \epsilon_U, \left\vert \bar \eta_U \right\vert) \  =  1 \, .
\end{align}

In single field inflation, the end-of-inflation conditions imposes a field value $\varphi = \varphi_0$ at~$N=0$. Using this value, it is possible to determine~$N(\varphi)$ through~\eqref{efolds}, which we may invert and substitute in the first order expressions for the cosmological observables~$r(\varphi)$,~$n_\mathcal{R}(\varphi)$~{\it etc}, leading to expressions entirely in terms of the number of $e$-folds~$N$. However, in the presence of $n$ evolving scalar fields, each value of $N = N(\varphi)$ describes an~$(n-1)$-dimensional \emph{isochrone surface} in field space, with $N=0$ corresponding to the end-of-inflation surface. We expect observable quantities to change as we select end-of-inflation field values on the~$N=0$ isochrone. This $N=0$ isochrone may be parameterized by a set of~$n-1$ parameters~$\lambda^I = (\lambda^1, \ldots, \lambda^{n-1})$. Thus, there are many possible inflationary trajectories and we require some criterion in order to select the observationally viable ones. In this respect, an important constraint is the normalization of the theoretical prediction for $P_\mathcal{R}$ to the observed scalar power spectrum.

Another question that naturally arises is how sensitive a given inflationary model is to the boundary conditions. Slightly changing the boundary conditions on the end-of-inflation isochrone might lead to a large change on the horizon crossing isochrone $N=N_*$, thereby greatly affecting the values of the observable quantities. Therefore, it is useful to introduce a parameter that quantifies the stability of inflationary trajectories. To this end, we first consider the space of possible trajectories, parameterized by the field values at the end of inflation as constrained by \eqref{maxgen}. We first consider an arbitrary isochrone at a given number of $e$-folds $N$, parameterized by $\lambda^I$ and described  by $\varphi^A = \varphi^A_N(\lambda)$. The geometry of this isochrone is encoded in the \emph{induced metric}~$[\Gamma_{IJ}]_N$, given by
\begin{align}
\label{indmetric}
[\Gamma_{IJ}]_N \ =\  G_{AB}  \; \left( \nabla_I \varphi^A_{N}\right) \;\left( \nabla_J \varphi^B_{N}\right)\; ,
\end{align}
where $I,J$ are indices which correspond to the vectors living on the tangent space of this particular $(n-1)$-dimensional surface. All frame-covariant quantities on this surface are generated using the induced metric $[\Gamma_{IJ}]_N$, which inherits the conformal weight and scaling dimension of $G_{AB}$. 

We may define the \emph{density of trajectories} $n  (N)$ for a neighbourhood~$(\lambda^I, \lambda^I + d\lambda^I)$ (which has a corresponding area element $dS_N$ on the isochrone) as follows:
\begin{align}
n (N)\ \equiv\ \frac{1}{S_N}\frac{d^{n-1} S_N}{d^{n-1} \lambda}\ .
\end{align}
Note that this definition of the density of trajectories ensures that the surface integral of~$n(N)$ over any isochrone $N$ is $1$. The area element $d^{n-1} S_N$ that corresponds to the neighbourhood~$(\lambda^I, \lambda^I + d\lambda^I)$ under consideration is given by
\begin{align}
\label{hyperdS}
d^{n-1} S_N = d^{n-1} \lambda \, \sqrt{\left\vert\det  [\Gamma_{IJ}]_N\right\vert} \ ,
\end{align}
leading to the following expression for the density of trajectories:
\begin{align}
\label{dot}
n (N)\ =\ \frac{\sqrt{\left\vert\det  [\Gamma_{IJ}]_N\right\vert}}{\int d^{n-1} \lambda \, \sqrt{\left\vert\det  [\Gamma_{IJ}]_N\right\vert}} \ .
\end{align}
We may call a  trajectory \emph{stable} at $N=N_2$ with respect to its boundary conditions at~$N=N_1$ if the density of trajectories increases as we move from $N_1$ to $N_2$. Similarly, an \emph{unstable} trajectory is one along which the density of trajectories decreases. Thus, we may define the \emph{sensitivity} parameter $Q(N_1,N_2)$ for a trajectory at the $N = N_2$ isochrone to the boundary conditions at the $N = N_1$ isochrone as follows:
\begin{align}
\label{relsens}
Q  (N_1,N_2)\   \equiv\ \frac{n  (N_2) }{n  (N_1) }\ ,
\end{align}
where $n(N_1)$ and $n(N_2)$ must be evaluated along the trajectory of interest. For our purposes, we are interested in the sensitivity of the field values at horizon crossing $N_2 = N_*$ to the boundary conditions at $N_1 = 0$, since it is the values of the fields at horizon exit that impact the inflationary observables. To this end, we define~$Q_*  \equiv Q (N_*, 0)$, observing that~$Q_* < 1$ corresponds to stable trajectories with respect to boundary conditions, whereas~$Q_* > 1$ corresponds to unstable trajectories. In the next section, we will consider specific two-field models of inflation and examine the stability of inflationary trajectories, as well as the aforementioned normalization criterion for singling out those trajectories that require the least degree of fine-tuning.

\section{Specific Models}
\label{specmod}

In this section, we apply the multifield frame-covariant formalism to two simple multifield models: (i) a two-field minimal model and (ii) a two-field nonminimal model inspired by Higgs inflation. We study the different trajectories admissible in each theory and select a particular trajectory by normalizing to the observable power spectrum $P^\text{obs}_\mathcal{R}$, presenting numerical predictions for the inflationary observables using the analytic results from Section~\ref{quantpert}. For both models, we take into account the entropy transfer and evaluate its effect on the inflationary observables. We conclude by outlining how our approach may be used to study~$F(\varphi, R)$ models by showing that they are equivalent to multifield scalar-curvature theories through the method of Lagrange multipliers.

\subsection{Minimal Two-Field Inflation}
\label{mininfl}

\begin{figure} 
  \centering
    \includegraphics{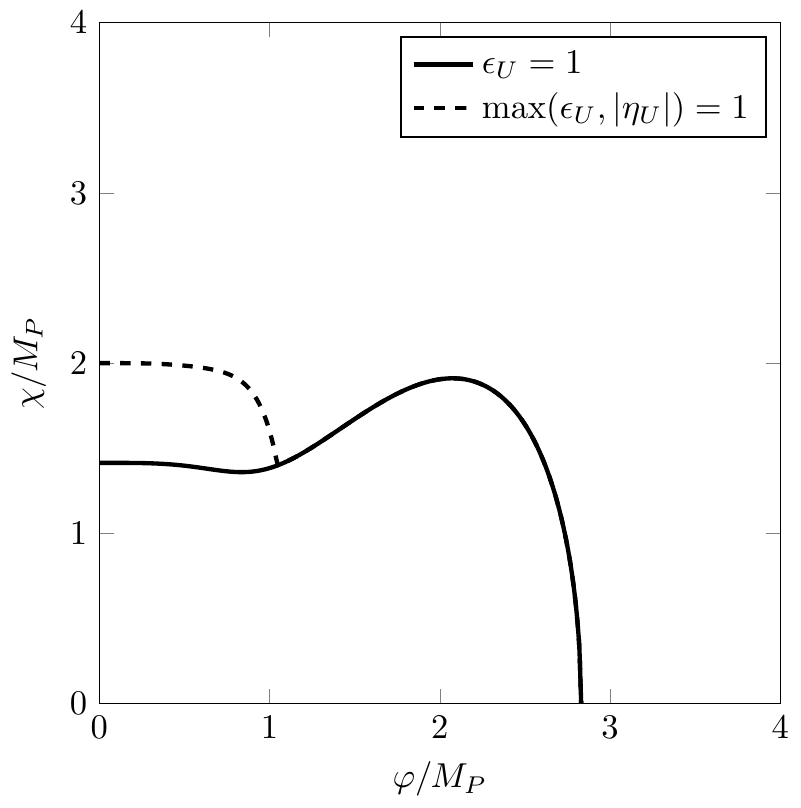}
  \caption{End-of-inflation curve for the minimal model with  $m^2/(\lambda M_P^2) = 1$.}
  \label{fig:endofinflation}
\end{figure}

\begin{figure} 
  \centering
  \hspace{-1em}
\includegraphics{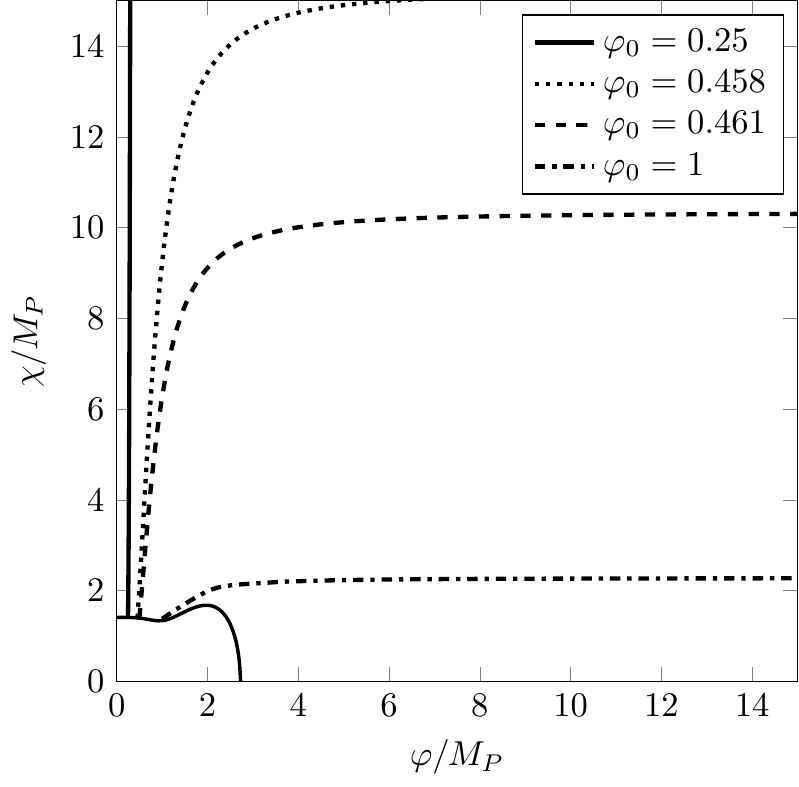}
 \includegraphics{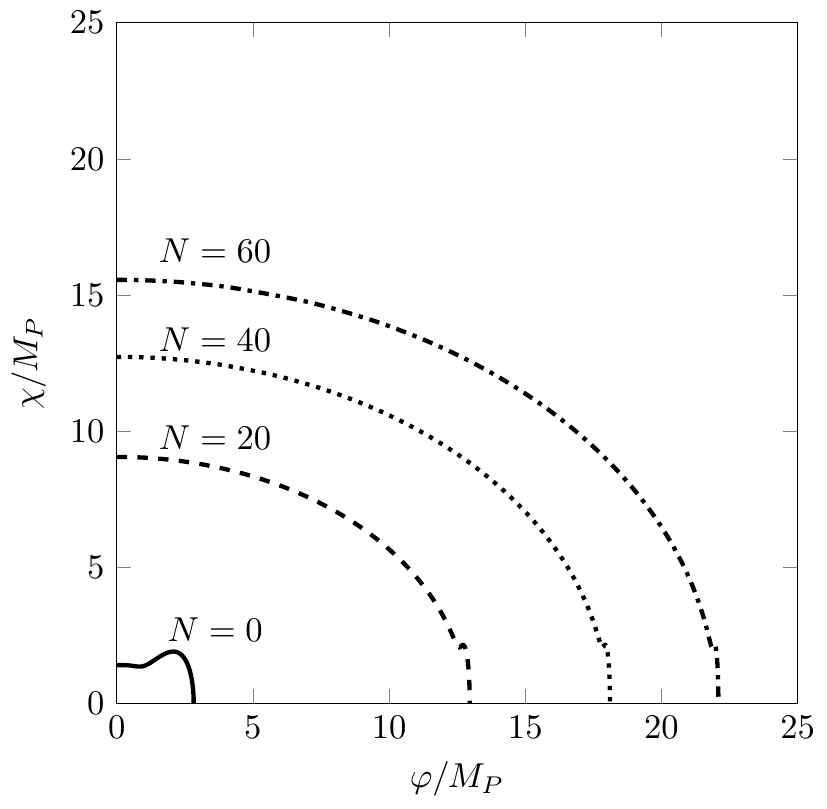}
  \caption{Field space trajectories and isochrone curves for the minimal model.}
  \label{fig:trajectories}
\end{figure}

\begin{figure} 
  \centering
\includegraphics{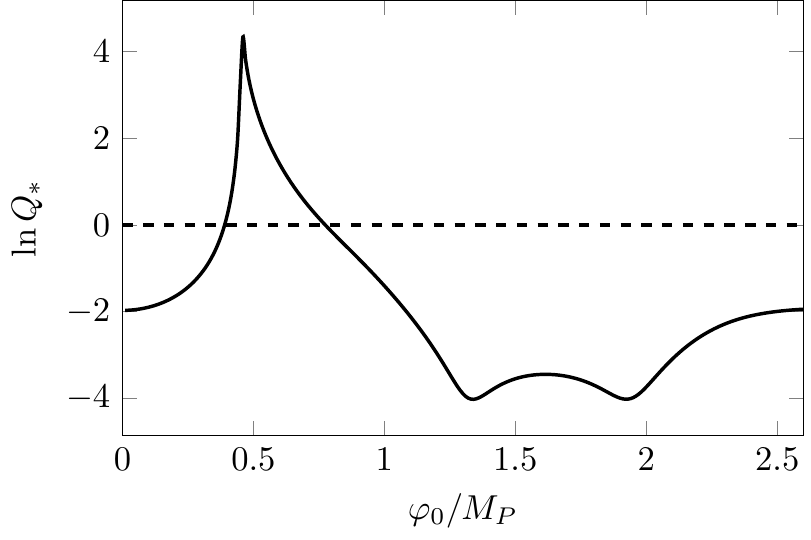}
  \caption{Sensitivity parameter $Q_*$  for the minimal model at $N=60$ to boundary conditions given by $\varphi_0$. The dashed line corresponds to $Q_* = 1$.}
  \label{fig:qMIN}
\end{figure}

We first examine a simple minimal two-field model, described by the Lagrangian
\begin{align}
\label{lagr1}
\mathcal{L} = -\frac{ M_P^2 R}{2} + \frac{1}{2} (\nabla \varphi)^2 +  \frac{1}{2} (\nabla \chi)^2 - \frac{\lambda\varphi^4 }{4}  - \frac{m^2 \chi^2}{2},
\end{align}
where $m$ is a mass parameter and $\lambda$ is the quartic coupling~\cite{GarciaBellido:1995qq}. This model is distinct from hybrid inflation \cite{Linde:1993cn}, where one field acts like a ``waterfall'' field. This is because both fields slowly roll down the inflationary potential and as such, both can act as inflaton fields. Employing the potential slow-roll parameters, we may calculate the general form of the observables in terms of $\varphi$ and $\chi$. Once the values of the fields $\varphi_*$ and $\chi_*$ are known at the time of horizon crossing for scales of interest, the values of all cosmological observables can be computed through \eqref{potobs}. In addition, we may specify the end-of-inflation condition by requiring $\bar \epsilon_U = 1$, which in turn implies
\begin{align}
\label{eoinfmin}
\frac{8 M_P^2 \left(\lambda ^2 \varphi_0^6+m^4 \chi_0^2\right)}{\left(\lambda  \varphi_0^4+2 m^2 \chi_0 ^2\right)^2} \ =\ 1\; .
\end{align}
Solving \eqref{eoinfmin} for $m^2/(\lambda M_P^2) = 1$ yields
\begin{align}
\label{chi0}
\chi_0 (\varphi_0) \ =\ \Big[\,1  - \frac{1}{2}(\varphi_0/M_P)^4 +\sqrt{1 -(\varphi_0/M_P)^4 + 2 (\varphi_0 /M_P) ^6 }\,\Big]^{1/2} M_P\;.
\end{align}
As can be seen in Figure~\ref{fig:endofinflation}, \eqref{chi0} defines the end-of-inflation contour $\bar\epsilon_U = 1$ in field space corresponding to $N=0$, for which $\varphi_0$ varies from $0$ to $\sqrt{8}M_P$.
As we choose different boundary conditions on this isochrone, the field trajectories change as shown in Figure~\ref{fig:trajectories}.
We further observe that there exists some critical region for $\varphi_0$ in which the trajectories are unstable. It is useful to parametrize the horizon crossing curve at $N = N_*$ in terms of $\varphi_0$:
\begin{align}
\chi_* = \chi_* (\varphi_0), \qquad \varphi_* = \varphi_* (\varphi_0).
\end{align}
Then, we may use \eqref{indmetric} in order to compute the values of the induced metrics $[\Gamma_{IJ}]_N$ on the two isochrones $N=0$ and $N=N_*$ and substitute them in \eqref{relsens}. In this case,  the sensitivity parameter~$Q_*$ becomes
\begin{align}
\label{qmin}
Q_*(\varphi_0)\
= \
\dfrac{
\sqrt{\left(\frac{d\varphi_*}{d\varphi_0}\right)^2 + \left(\frac{d\chi_*}{d\varphi_0}\right)^2}
\bigg/
{\displaystyle\int}^{\varphi_{0\text{max}}}_0 d \varphi_0   \, 
 \sqrt{\left(\frac{d\varphi_*}{d\varphi_0}\right)^2 + \left(\frac{d\chi_*}{d\varphi_0}\right)^2}
 }
{
\sqrt{1+ \left(\frac{d\chi_0}{d\varphi_0}\right)^2}
\bigg/
{\displaystyle\int}^{\varphi_{0\text{max}}}_0 d \varphi_0   \, 
 \sqrt{1+ \left(\frac{d\chi_0}{d\varphi_0}\right)^2}
}\ ,
\end{align}
with $\varphi_{0\text{max}} = \sqrt{8} M_P$.  In Figure~\ref{fig:qMIN} we plot $\ln Q_*$ as a function of $\varphi_0$.
We observe that trajectories with~$\varphi_0/M_P \in [0.391,0.775]$  are unstable and the sensitivity parameter $Q_*$ is maximized around the trajectory for which $\varphi_0 = \varphi_\text{crit} \equiv 0.458 \, M_P$.
 
\begin{figure} 
  \centering
  \hspace{-1em}
\includegraphics{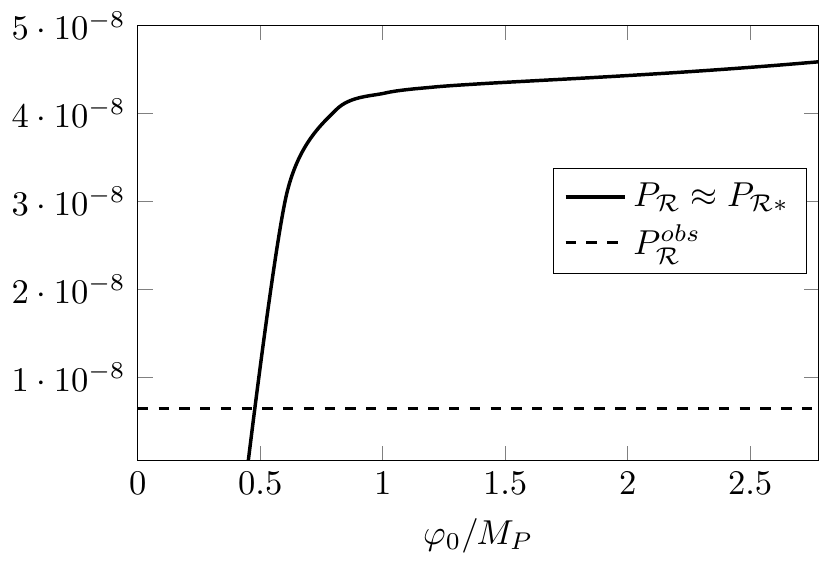}
\includegraphics{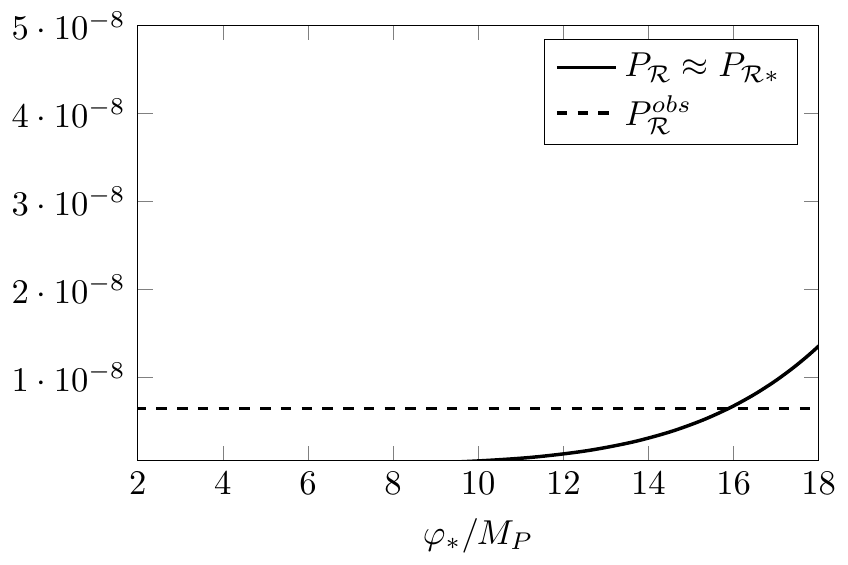}
  \caption{Power spectrum normalization for the minimal model with $\lambda  = 10^{-12}$ and $m/M_P = 10^{-6}$ at~$N=60$ for different boundary conditions in terms of $\varphi_0$ and the corresponding horizon crossing values $\varphi_*$. Solid lines correspond to the theoretical predictions while the horizontal line corresponds to the observed power spectrum $P^\text{obs}_\mathcal{R}$ given in \eqref{pobs}.}
  \label{fig:PRMIN}
\end{figure}

\begin{figure}
\hspace*{-0.55em}
\includegraphics{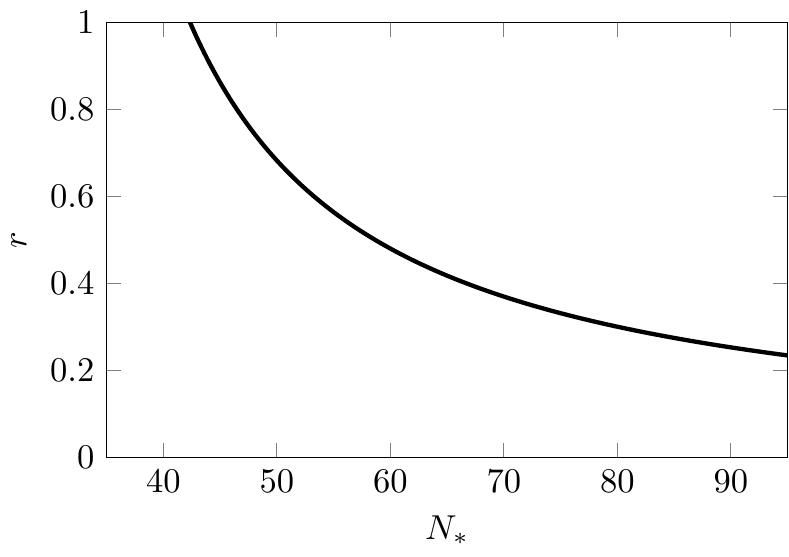} 
\hspace{1.8em}
 \includegraphics{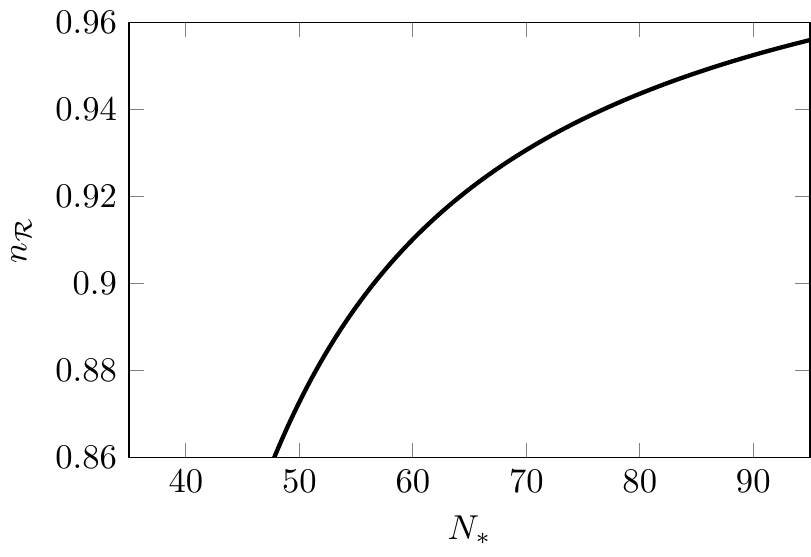}   
\\[2mm]
\hspace*{-2em}
 \includegraphics{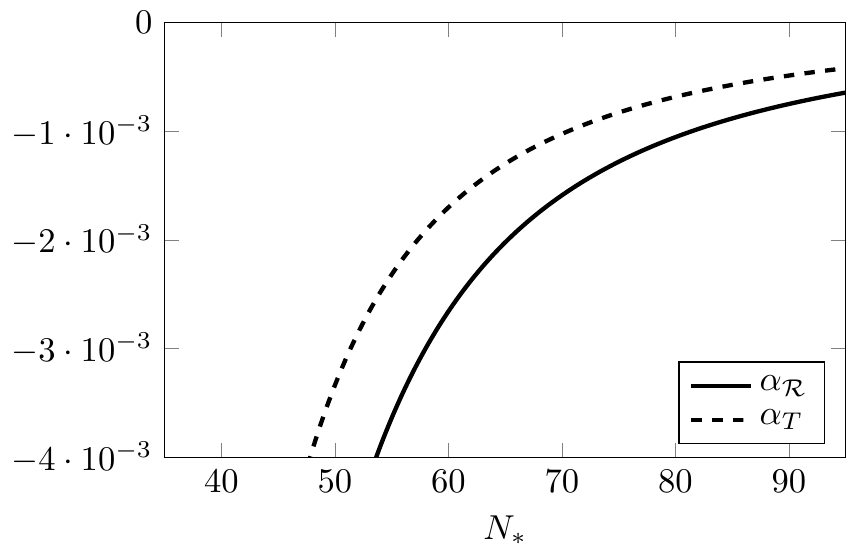}
 \includegraphics{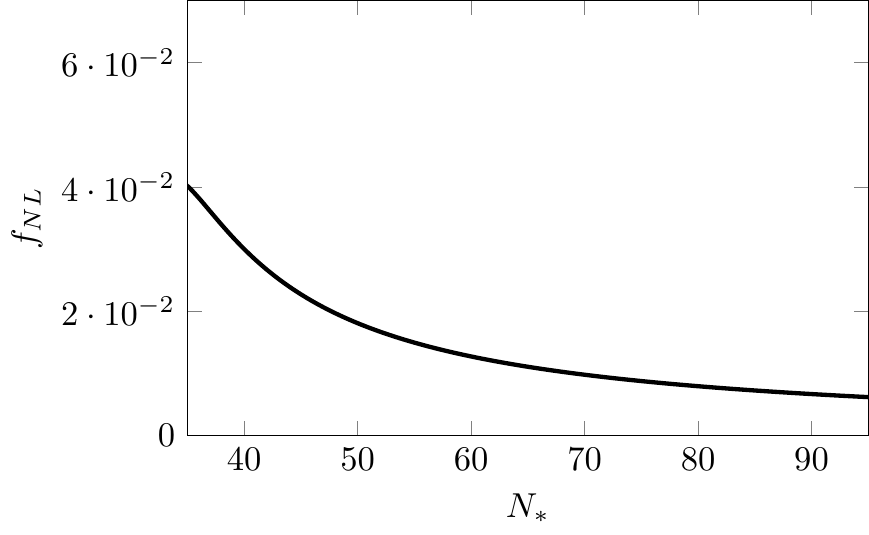}
  \\[2mm]
 \hspace*{9em}
 \includegraphics{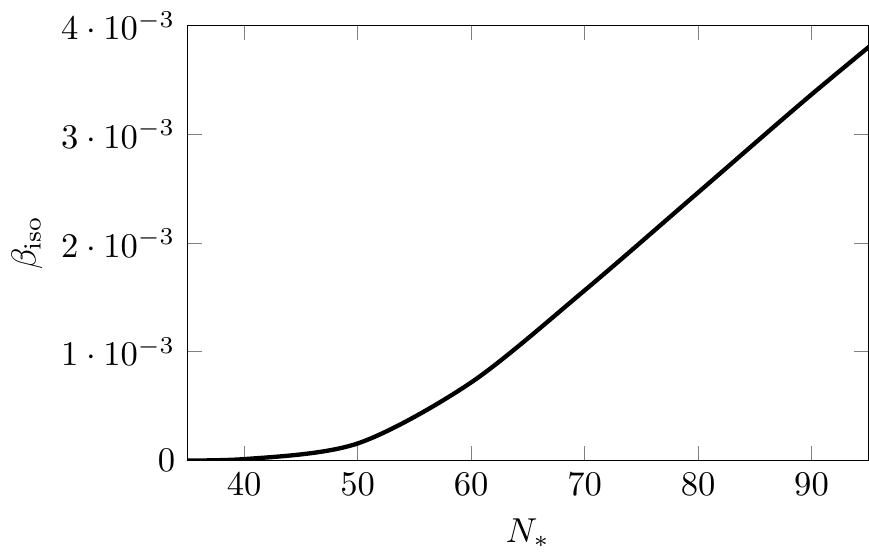}
  \caption{Predictions for the inflationary quantities $r , n_\mathcal{R}, \alpha_\mathcal{R}, \alpha_T$, $f_{NL}$ and $\beta_{\rm iso}$ in the minimal model for boundary condition given by $\varphi_0/M_P= 0.495$.}
  \label{fig:observablesMIN}
\end{figure}

We now single out a particular trajectory by matching the value of $P_\mathcal{R}$ to the observed scalar power spectrum at the $68 \%$ confidence level \cite{Ade:2015lrj}
\begin{align}
\label{pobs}
P^\text{obs}_\mathcal{R} = (6.41 \pm 0.18) \times 10^{-9}.
 \end{align}
We may numerically calculate $P_\mathcal{R}$ for the model at hand for different values of $\varphi_0$. We choose~$\lambda=10^{-12}$ and~$m=10^{-6} M_P$, and display the predicted power spectra $P_\mathcal{R}$ at the end of inflation and $P_{\mathcal{R}*}$ at horizon crossing as functions of $\varphi_0$ in Figure~\ref{fig:PRMIN}.
We find that the entropy transfer has a negligible effect on the boundary condition~$\varphi_0/M_P= 0.495\pm 0.001$ compatible with $P^\text{obs}_\mathcal{R}$.  The power spectrum normalization $P_\mathcal{R} = P^\text{obs}_\mathcal{R} $ may be used to relate~$\varphi_0$ to the parameters of the theory and further restrict the parameter space. In this model, we have chosen particular values for the parameters, leading to a unique prediction for~$\varphi_0$. Hence, after numerically solving for $\varphi(N)$ and $\chi(N)$ with $\varphi(0) = \varphi_0$ and $\chi(0) = \chi_0 (\varphi_0)$, we may substitute the resulting solutions in the expressions for the observables obtained through \eqref{potobs} in order to plot the $r, n_\mathcal{R}, \alpha_\mathcal{R}, \alpha_T, f_{NL}$, and $\beta_{\rm iso}$ as functions of $N$ for different values of $\varphi_0$ in Figure~\ref{fig:observablesMIN}.
\begin{figure}
\hspace*{-2.55em}
\includegraphics{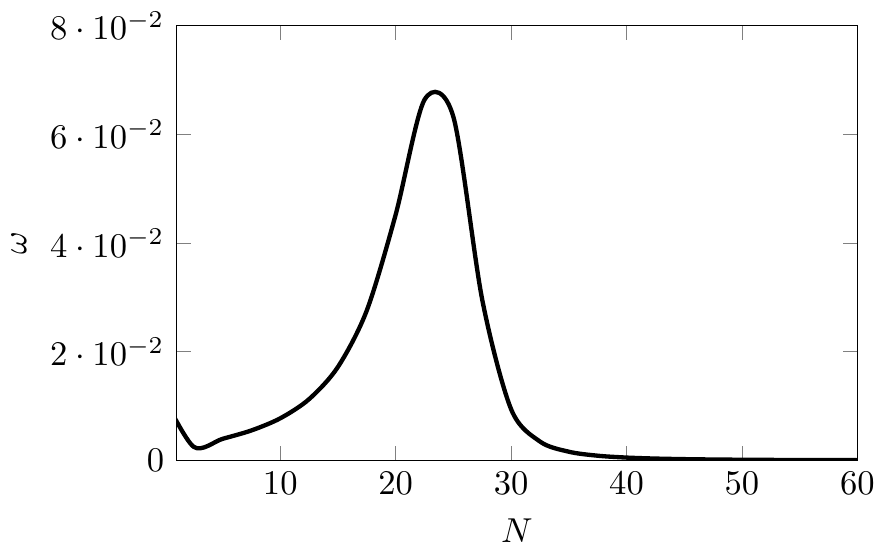} 
\hspace{1.8em}
 \includegraphics{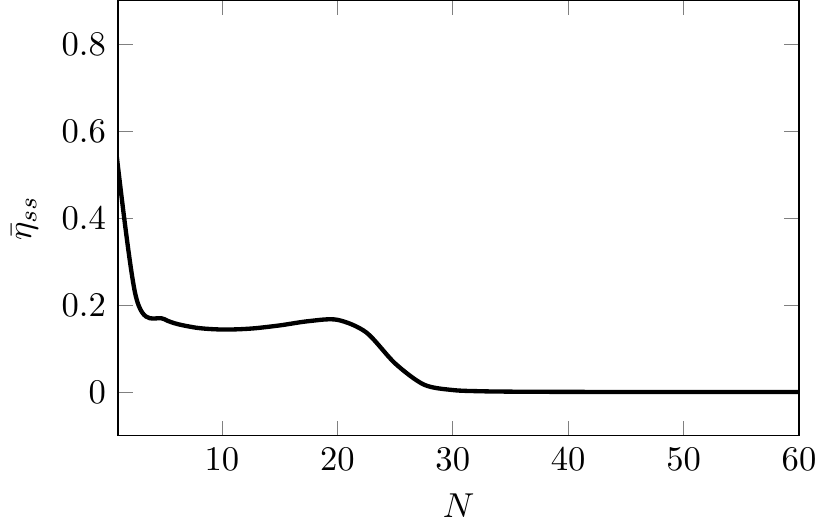} 
    \caption{Evolution of $\omega$ and $\bar \eta_{ss}$ along the inflationary trajectory with $\varphi_0 = 0.495$ for the minimal two-field model. \label{fig:turnrateMIN} }
\end{figure}

We may summarize the predictions of this model in Table \ref{tab:observablesmin} and compare our predictions to the currently observed values. We see from Figure \ref{fig:turnrateMIN} that the turn rate achieves its maximum at around 20 e-folds, but the generation of isocurvature modes is suppressed since~$\bar \eta_{ss}$ does not achieve a negative value throughout the inflationary trajectory. Therefore, the entropy generation in this model are negligible, as reflected in the smallness of $\beta_{\rm iso}$ (see last panel in Figure \ref{fig:observablesMIN}). We also note that the values of $r$ and $n_\mathcal{R}$ lie far outside the PLANCK bounds.  For this reason, we extend this model by adding a \emph{nonminimal parameter} to the Lagrangian in the next subsection.
 \begin{table}[!h]
\centering

\begin{tabular}{ l  l  l  l}
  &   $\hphantom{-}\varphi_0/M_P= 0.495$		& \hspace{0.5em} PLANCK 2015 \\	
  \hline
  $r$ 				& $\hphantom{-}0.501 $ 					& \hspace{0.5em}$\le 0.12 \ (  95 \%   \text{ CL}) $ \\ 
  $n_\mathcal{R}$		& $\hphantom{-}0.906 $ 					& $\hphantom{-}0.968 \pm 0.006 \ (  68 \%   \text{ CL})   $  \\ 
  $\alpha_\mathcal{R}$  	& $-0.00288 $  						& $-0.003 \pm 0.008  \ (  68 \%   \text{ CL})$     \\
  $\alpha_T$ 			& $-0.0019 $ 						& $-0.000167 \pm  0.000167 \ (  68 \%   \text{ CL})$ \\ 
  $f_{NL}$			& $\hphantom{-}0.0129$					& $\hphantom{-}0.8 \pm 5.0 \  (   68 \%   \text{ CL})$ \\
  $\beta_{\rm iso}$		& $\hphantom{-}0.000717$					&  \hspace{0.5em}$\le 0.08 \ {\rm (CDI)}, 0.27 \ {\rm (NDI)}, 0.18 \ {\rm (NVI)}   \ (  95 \%   \text{ CL}) $
\end{tabular}
\caption{Observable inflationary quantities for the minimal two-field model at $N=60$.
  Note that the running of the tensor spectral index $\alpha_T$ is not quoted in \cite{Ade:2015lrj}, as no tensor modes were measured by PLANCK. It~is derived from the consistency relation \eqref{2ndconseq} with transfer angle $\Theta = 0$, and serves as a constraint on a possible future measurement of $\alpha_T$, in the slow-roll approximation. The parameter $\beta_{\rm iso}$ is constrained by assuming different non-decaying isocurvature modes: (i) the cold dark matter density isocurvature mode (CDI), (ii) the neutrino density mode (NDI), and (iii) the neutrino velocity mode (NVI).
      \label{tab:observablesmin}}
\end{table}

\subsection{Nonminimal Two-Field Inflation}

 A possible extension to the model described in the previous subsection is to introduce a coupling between one of the fields and the scalar curvature $R$ as inspired by Higgs inflation~\cite{Bezrukov:2007ep}. The Lagrangian of this model is given by
\begin{align}
\label{lagr2}
\mathcal{L} \ = \ -\frac{(M_P^2+ \xi \varphi^2) R}{2} + \frac{1}{2} (\nabla \varphi)^2 +  \frac{1}{2} (\nabla \chi)^2 - \frac{\lambda(\varphi^2-v^2)^2 }{4}  - \frac{m^2 \chi^2}{2}\ ,
\end{align}
where we assume that the VEV $v$ is negligible in the inflationary regime $\varphi \sim M_P$. Unlike in the minimal scenario discussed in Subsection~\ref{mininfl}, we deviate from $m^2/(\lambda M_P^2) = 1$ by choosing $m = 5.6 \times 10^{-6} \, M_P$ and $\lambda = 10^{-12}$, as well as $\xi = 0.01$ for the nonminimal parameter. This model features a curved field space, as we can see by calculating the field space Ricci scalar:
\begin{align}
S\ =\ \frac{2 \xi  \left[M_P^4-\xi ^2 (1 + 6 \xi)  \varphi^4\right]}{\left[M_P^2+\xi  (1+ 6 \xi) \varphi^2\right]^2}\ .
\end{align}
Assuming $\xi >0$, we observe that the curvature is positive, $S \approx 2\xi$  in the regime where $\varphi$ is small, given by $\varphi/M_P \ll 1/\sqrt{\xi}$. Conversely, in the regime where $\varphi$ is large, the curvature becomes negative, $S \approx - 2\xi/(1+6\xi)$. For $\xi$ of order $10^3$, such as  in \cite{Schutz:2013fua}, the trajectory is in the region of the field space with negative curvature for most of inflation, which contributes greatly towards the generation of isocurvature perturbations. In our case with $\xi = 0.01$, the curvature is positive but small, which means that isocurvature perturbations will not necessarily be generated towards the end of inflation.

\begin{figure}
  \centering
    \includegraphics{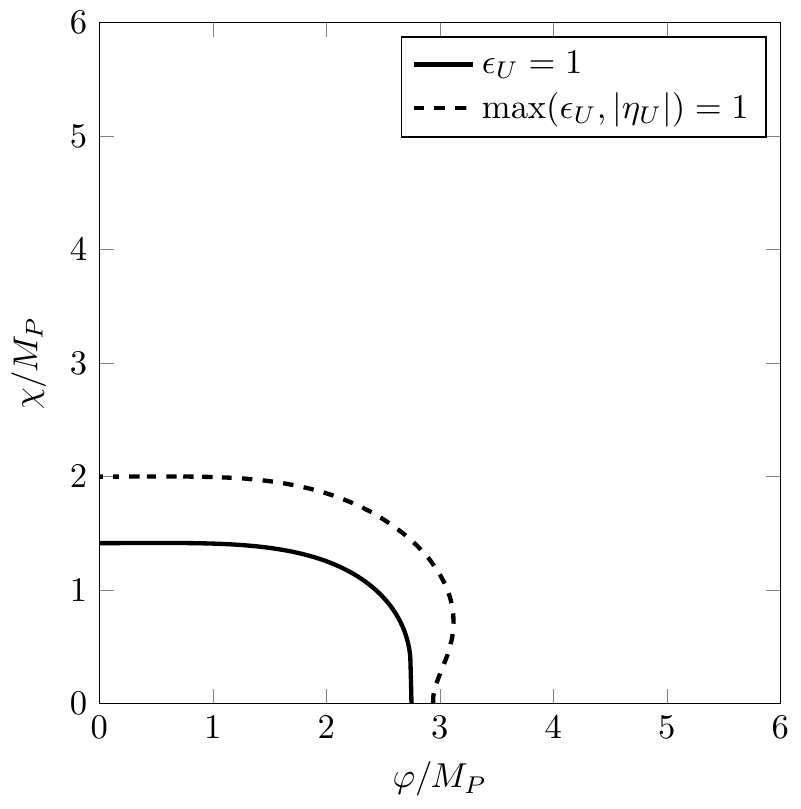}
  \caption{End-of-inflation curve for the nonminimal model with $m = 5.6 \times 10^{-6}\, M_P$, $\lambda = 10^{-12}$, and~$\xi = 0.01$.}
  \label{fig:endofinflationNM}
\end{figure}

\begin{figure}
  \centering
  \hspace{-1em}
    \includegraphics{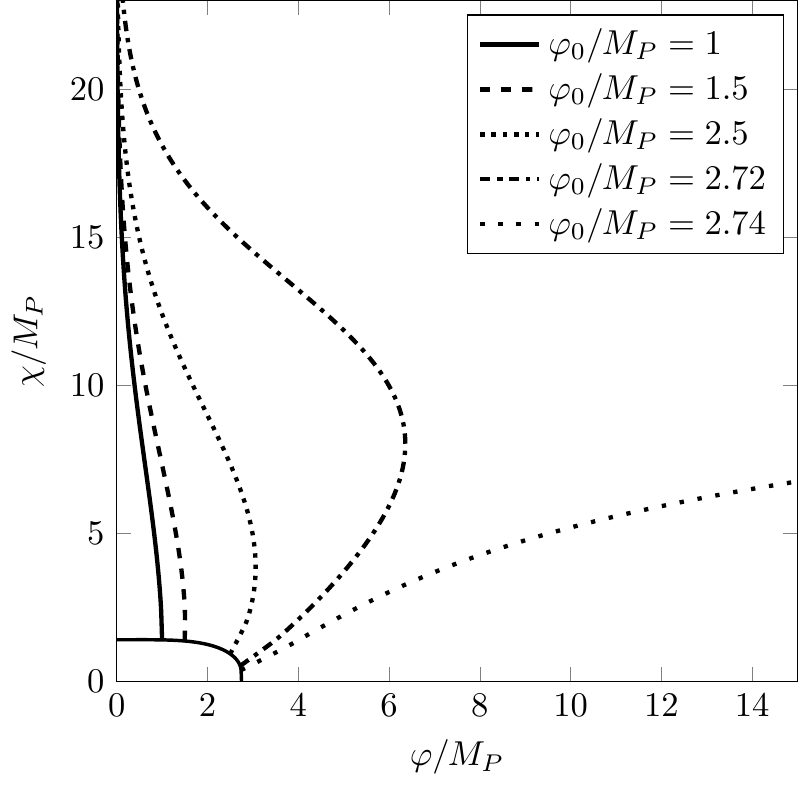}
    \includegraphics{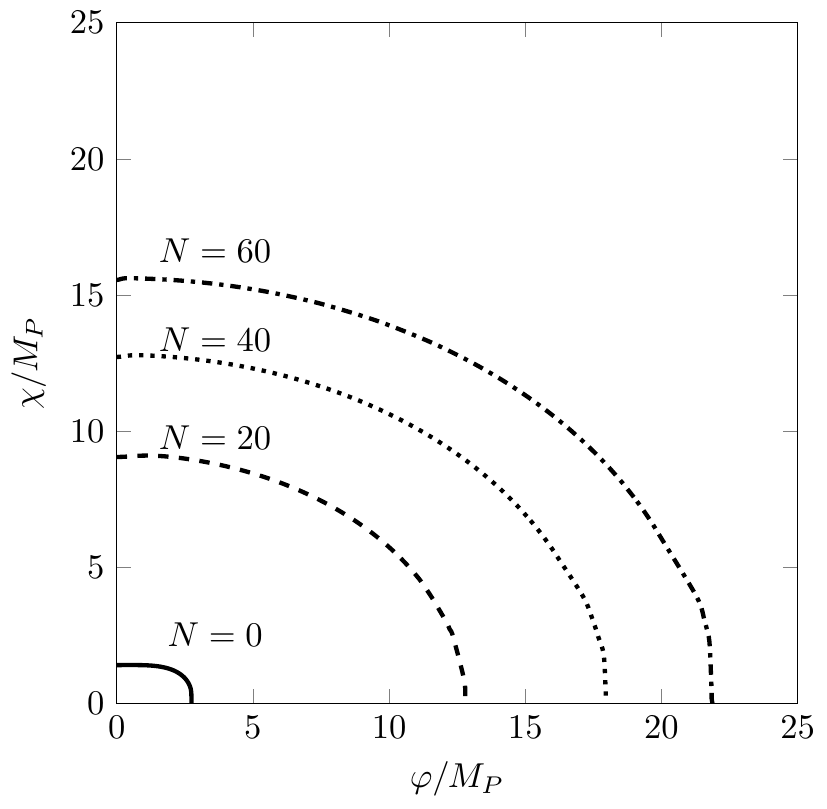}
  \caption{Field space trajectories
and isochrone curves
  for the nonminimal model.}
  \label{fig:phichiplotNM}
\end{figure}

\begin{figure} 
 \hspace{9.75em}
 \includegraphics{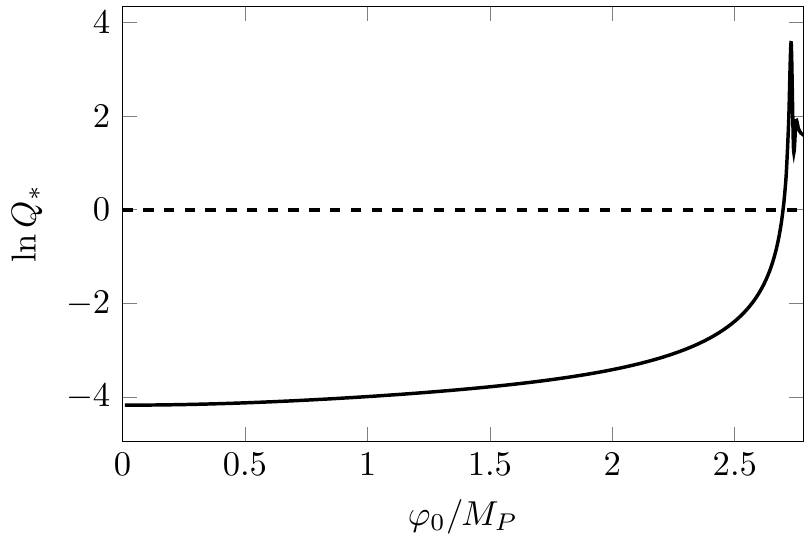}
  \caption{Sensitivity parameter $Q_*$ for the nonminimal model to boundary conditions given by $\varphi_0$. The dashed line corresponds to $Q_* = 1$.
  }
  \label{fig:qNM}
\end{figure}

Proceeding as in the minimal model, we show the end-of-inflation contour in Figure~\ref{fig:endofinflationNM}.
In~Figure~\ref{fig:phichiplotNM} we display the field space trajectories and in Figure \ref{fig:qNM}  the sensitivity parameter~$Q_*$ for all possible trajectories on the boundary conditions.
Specifically, the parameter $Q_*$ may be cast into the form
\begin{align}
\label{qNM}
\hspace{-1em}
Q_*(\varphi_0)\
\equiv \
\dfrac{
\sqrt{G^*_{\varphi\varphi } \left(\frac{d\varphi_*}{d\varphi_0}\right)^2 + G^*_{\chi\chi}\left(\frac{d\chi_*}{d\varphi_0}\right)^2}
\bigg/
{\displaystyle\int}^{\varphi_{0\text{max}}}_0 d \varphi_0   \, 
 \sqrt{G^*_{\varphi\varphi } \left(\frac{d\varphi_*}{d\varphi_0}\right)^2 +G^*_{\chi\chi } \left(\frac{d\chi_*}{d\varphi_0}\right)^2}
 }
{
\sqrt{G^0_{\varphi\varphi }+ G^0_{\chi\chi} \left(\frac{d\chi_0}{d\varphi_0}\right)^2}
\bigg/
{\displaystyle\int}^{\varphi_{0\text{max}}}_0 d \varphi_0   \, 
 \sqrt{G^0_{\varphi\varphi }+  G^0_{\chi\chi} \left(\frac{d\chi_0}{d\varphi_0}\right)^2}
}\ ,
\end{align}
where $G^*_{AB}$ and $G^0_{AB}$ correspond to the values of the field space metric on the two different isochrone curves $N = N_*$ and $N = 0$, respectively, and $\varphi_{0\text{max}}/M_P \approx 2.742$.
We observe that~$Q_*$ is small for most values of the boundary conditions. This agrees well with the left panel of Figure \ref{fig:phichiplotNM}, where trajectories with different boundary conditions converge. The critical value is given by $\varphi_\text{crit}/M_P = 2.694$, close to the end of the $N=0$ isochrone.  

\begin{figure} 
  \centering
 \includegraphics{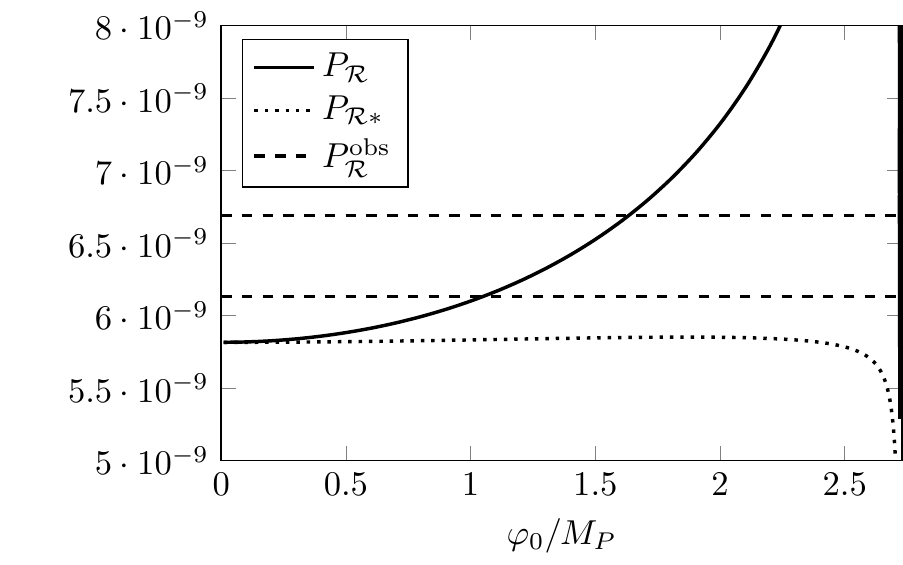}
  \hspace*{0.5em}
 \includegraphics{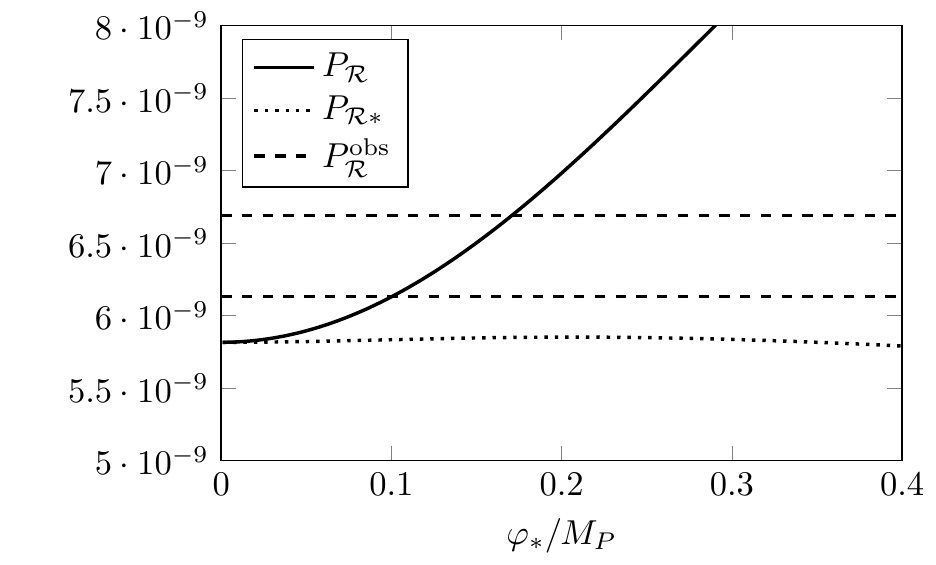}
  \caption{Power spectrum normalization for the nonminimal model with $m =5.6 \times 10^{-6} M_P, \lambda = 10^{-12}$, and~$\xi = 0.01$ for different boundary conditions in terms of $\varphi_0$ and the corresponding horizon crossing values~$\varphi_*$. Solid lines correspond to the theoretical predictions while the horizontal dashed lines correspond to the allowed band for the observed power spectrum $P^\text{obs}_\mathcal{R}$ given in \eqref{pobs}.}
  \label{fig:PRNM}
\end{figure}

In Figure~\ref{fig:PRNM}, we give predictions for the values of the power spectrum~$P_\mathcal{R}$.
From the left panel in Figure \ref{fig:phichiplotNM}, we see that an increasing boundary value $\varphi_0$  up to $\varphi_0/M_P \approx 2.73$ corresponds to a progressively sharper turn in field space.
The effects of entropy transfer have been studied in more detail in Figure~\ref{fig:PRNM}. Specifically, solid lines correspond to the full computation of $P_\mathcal{R}$, where the effect of entropy transfer is included. Instead, the dotted lines give the predictions for $P_{\mathcal{R}*}$, in which entropy transfer effects have been ignored. Observationally viable values for the boundary condition belong to the interval~$\varphi_0/M_P \in (1.048,1.634)$, as well as to~$\varphi_0/M_P = 2.72$. For the interval~$(1.048,1.634)$, the field value~$\varphi_0/M_P= 1.391$  corresponds to the mean value~$P_\mathcal{R}^\text{obs} = 6.41\times 10^{-9}$. We observe that the predicted value for~$P_\mathcal{R}$ is more sensitive to the value of $\varphi_0$ than in the minimal model.  

\begin{figure}
  \hspace*{-1.1em}
 \includegraphics{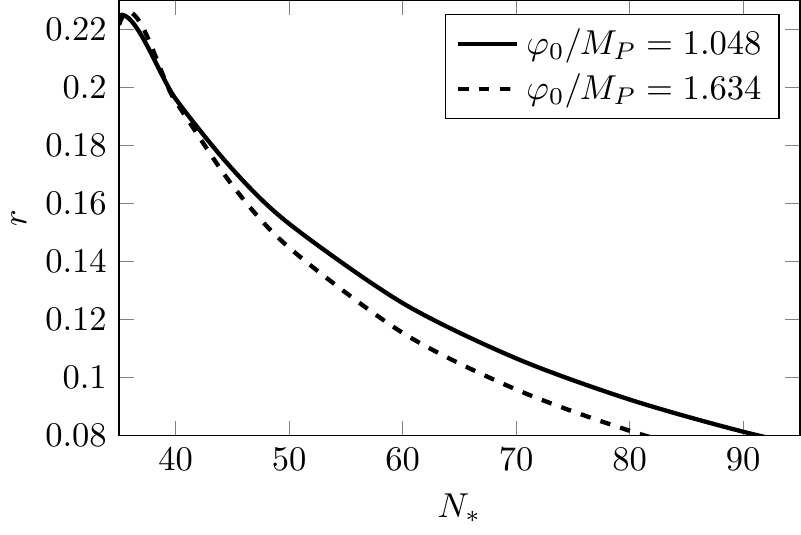}   
 \hspace*{2.2em}
 \includegraphics{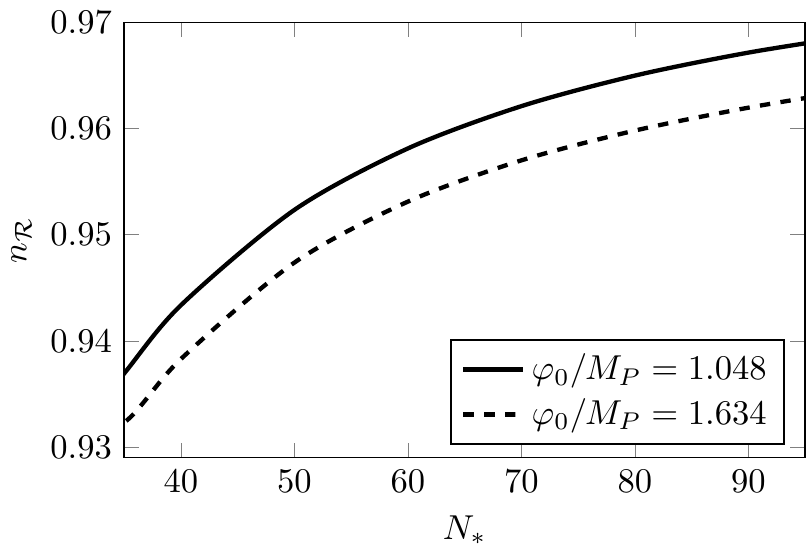} 
\\
\hspace*{-2.85em}
 \includegraphics{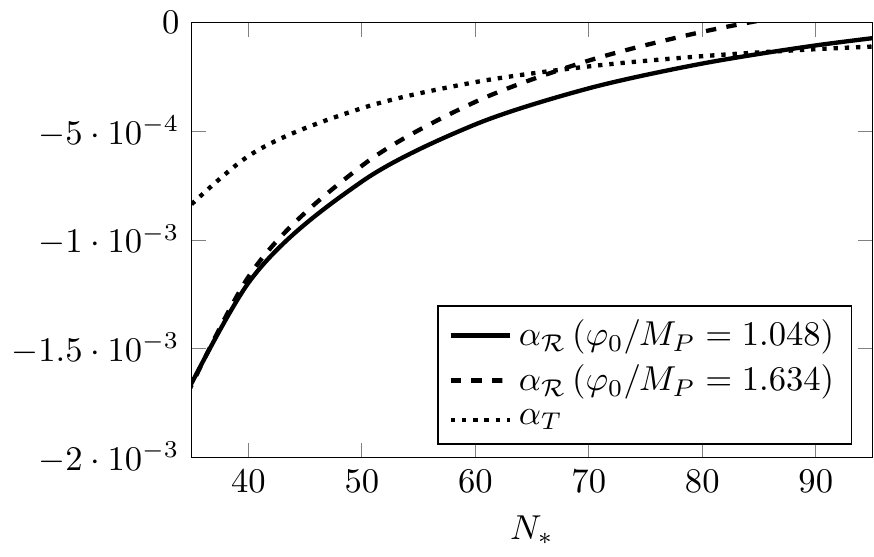}
 \includegraphics{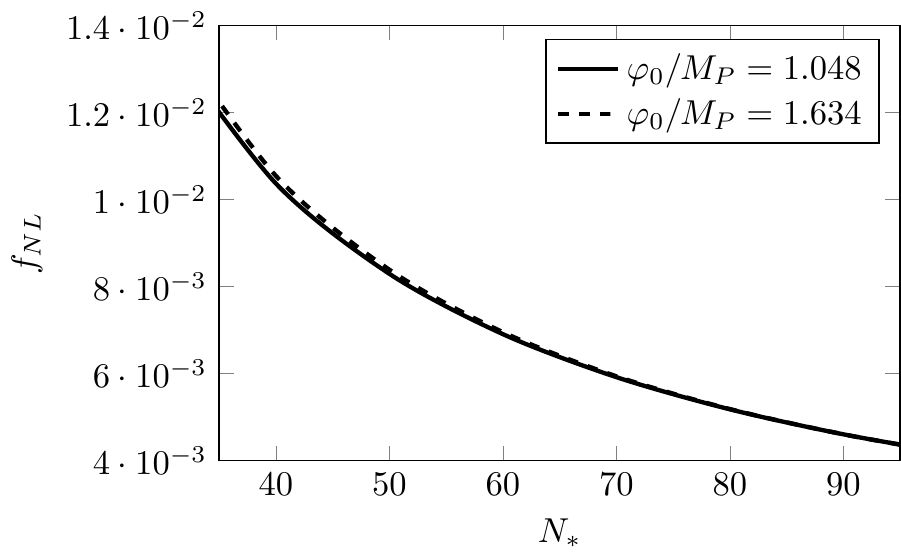}  
 \\
 \hspace*{9em}
 \includegraphics{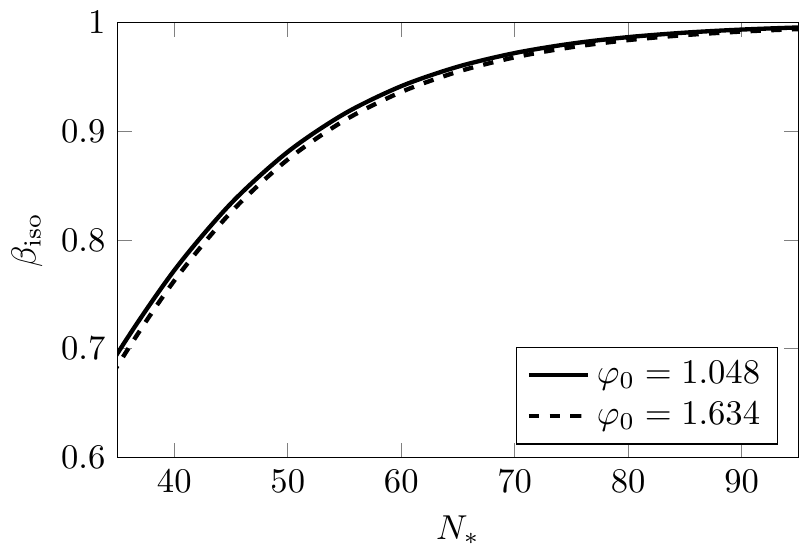}
  \caption{Predictions for the inflationary quantities $r , n_\mathcal{R}, \alpha_\mathcal{R}, \alpha_T$, $f_{NL}$ and $\beta_{\rm iso}$ in the nonminimal model for boundary conditions admissible under normalization of $P_\mathcal{R}$ to the observed power spectrum~$P^\text{obs}_\mathcal{R}$.}
  \label{fig:observablesNM}
\end{figure}

In Figure \ref{fig:observablesNM}, we show the dependence of the observables $r, n_\mathcal{R}, \alpha_\mathcal{R}, \alpha_T, f_{NL}$, and $\beta_{\rm iso}$ on the number of e-folds~$N$.  Taking $N = 60$ to be the point of horizon exit for the largest cosmological scales, we summarize our results in Table \ref{tab:observables} for the nonminimal two-field model.
We focus on the interval $\varphi_0/M_P \in  (1.048,1.634)$, as the trajectory for~$\varphi_0/M_P = 2.72$ generates predictions similar to~$\lambda \varphi^4$ inflation, which is not observationally viable.
\begin{table} 
\centering

\begin{tabular}{ l  l  l  l}
               			& \quad $ \varphi_0/M_P = 1.391^{+0.243}_{-0.343}$	&\hspace{0.5em}  PLANCK 2015  \\	
\hline	
$r$ 				&  $\hphantom{-}0.1204^{+0.0053}_{-0.0049}$   		&\hspace{0.5em}$\le 0.12 \  (   95 \%   \text{ CL})$ \\ 
$n_\mathcal{R}$		&  $\hphantom{-}0.955^{+0.005}_{-0.002} $		& $\hphantom{-}0.968 \pm 0.006 \  (   68 \%   \text{ CL}) $  \\ 
$\alpha_\mathcal{R}$  	&  $-0.0004^{+0.00005}_{-0.00006}$				& $-0.008 \pm 0.008 \  (   68 \%   \text{ CL})  $    \\
$\alpha_T$ 			&  $-0.000276^{+0.000003}_{-0.000003}$			& $-0.000155 \pm  0.00016 \  (   68 \%   \text{ CL}) $    \\ 
$f_{NL}$			&  $\hphantom{-}0.0693^{+0.00003}_{-0.00002}$	 	& $\hphantom{-}0.8 \pm 5.0  \  (   68 \%   \text{ CL})  $ \\
$\beta_{\rm iso}$		& $\hphantom{-}0.939^{+0.003}_{-0.003}$			&  \hspace{0.5em}$\le 0.08 \ {\rm (CDI)}, 0.27 \ {\rm (NDI)}, 0.18 \ {\rm (NVI)}   \ (  95 \%   \text{ CL}) $
\end{tabular}
\caption{Observable inflationary quantities for~the nonminimal model at $N=60$.  The limits on these quantities from 2015 PLANCK data \cite{Ade:2015lrj} are the same as in Table~\ref{tab:observablesmin}.
\label{tab:observables}}
\end{table}
We  note that this model agrees well with current observations within their uncertainties 
at the $2\sigma$ level displayed in Table \eqref{tab:observables}, except for $\beta_{\rm iso}$, which could be close to 1 (see the last panel in Figure \ref{fig:observablesNM}). However, the observables for our model were calculated at the end of inflation, not at the present epoch. As such, we have not taken into account reheating effects, which may cause the isocurvature perturbations to decay~\cite{Huston:2013kgl}. This would cause the empirically-determined value of $\beta_{\rm iso}$ to be much higher at the end of inflation, thus bringing our predictions within observational bounds. 
\begin{figure}
\hspace*{-3em}
\includegraphics{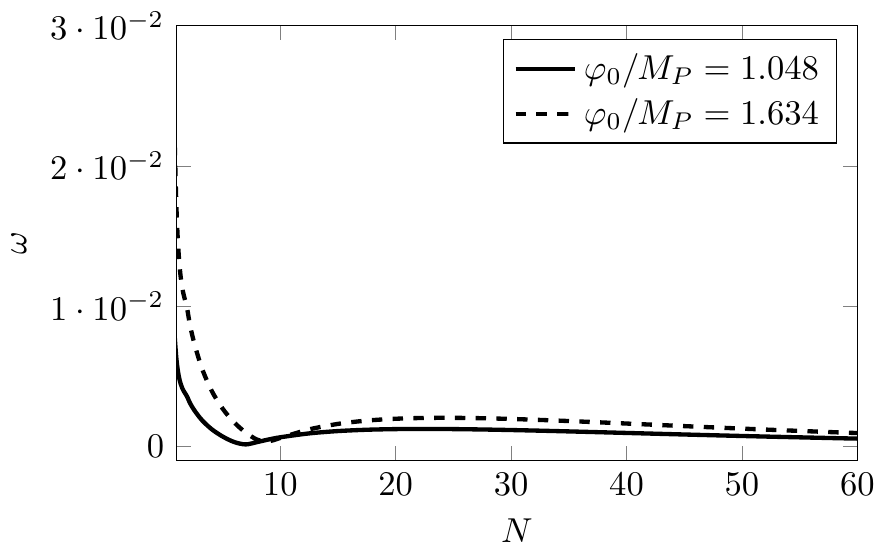} 
\hspace{1em}
 \includegraphics{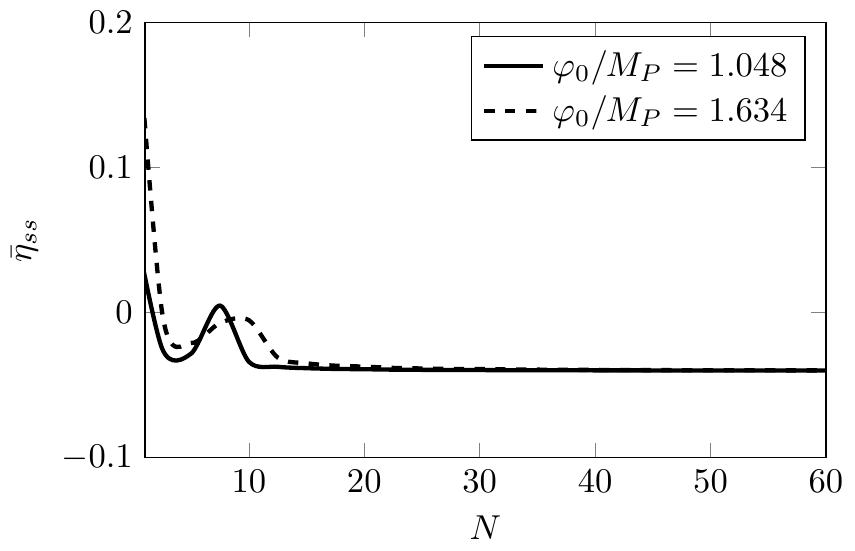} 
\caption{  \label{fig:turnrate2}
Evolution of $\omega$ and $\bar \eta_{ss}$ along the observationally viable inflationary trajectories for the nonminimal two-field model.}
\end{figure}

Notice that even the addition of a small nonminimal parameter $\xi$ modifies the cosmological observables by a significant amount and amplifies the effects of the entropy transfer on $P_\mathcal{R}$. The turn rate $\omega$ in this model does not vary considerably from the minimal case, as calculated from \eqref{turnrateapprox}. To lowest order in $\xi  \varphi ^2/M_P^2$, the turn rate $\omega$ is given by
 \begin{align}
 \omega\ =\  \frac{\varphi  \left(M_P^2 \xi -4 \xi ^2 \chi ^2\right)}{M_P^2    \chi}\ . 
\end{align}
We see from Figure \ref{fig:turnrate2} that $\omega$ is of order $10^{-3}$ throughout inflation, i.e.~one order of magnitude smaller than the one found in the minimal model. However, the amplification of isocurvature modes in this model is much larger than in the minimal model.  According to our discussion in Subsection~\ref{subsec:twofield}, this is because the parameter $\bar \eta_{ss}$ is negative for most of the two inflationary trajectories, as seen in Figure \ref{fig:turnrate2}. The effect of the positive field-space Ricci scalar $S$ on this model is negligible, since $S \approx 2\xi = 0.02$ is relatively small and also multiplied by $\bar\epsilon_H \ll 1$ in~\eqref{eta2}, giving rise to a suppressed contribution to $\bar \eta_{ss}$. Towards the end of inflation only, we have $\bar\epsilon_H \sim 1$ and a positive $S$ could suppress entropy production. But, 
the final stage of the inflationary era is too brief to affect the growth of isocurvature perturbations. This is in contrast to other models with non-minimal couplings in which entropy perturbations are amplified towards the end of inflation~\cite{Kaiser:2013sna}.  Thus, in our nonminimal model, the amplification of entropy is mainly driven by a concave potential $U$ in the isocurvature direction, whereas the turn rate $\omega$, albeit small, is sufficiently large to cause sizeable entropy transfer.

\subsection{$F(\varphi,R)$ Models}
Our approach can be straightforwardly extended to apply to a more general class of models, such as $F(\varphi,R)$ theories. Let us consider an $F(\varphi,R)$ theory given by the action
 \begin{align}
\label{fphiRact}
S = \int  d^4 x\,  \sqrt{-g}  \, \left[-\frac{F(\varphi,R)}{2}  + \frac{k_{AB}}{2} g^{\mu\nu }(\nabla_\mu \varphi^A) (\nabla_\nu \varphi^B) - V(\varphi) \right]\,.
\end{align}
We assume that the function $F(\varphi,R)$ must reduce to $M_P^2 R$ at the vacuum expectation value~$\varphi_{VEV}$ induced by $V(\varphi)$ such that the Einstein gravity limit is reached. Our goal is to write the action~\eqref{fphiRact} in terms of a multifield theory such that our formalism may be applied. The standard way to achieve this is to introduce a non-dynamical auxiliary degree of freedom $\Phi$ as a Lagrange multiplier and rewrite the action in \eqref{fphiRact} as
\begin{equation}
 \label{Raction}
 \begin{aligned}
S = \int  d^4 x\,  \sqrt{-g}  \, \bigg\{\!
-\frac{ 1}{2} \bigg[ F(\varphi, \Phi)  &+  F(\varphi,\Phi)_{,\Phi}\big(R-\Phi\big) \bigg]
\\
&+ \frac{k_{AB}}{2} g^{\mu\nu }(\nabla_\mu \varphi^A) (\nabla_\nu \varphi^B) - V(\varphi) \bigg\}\, .
\end{aligned}
\end{equation}
Varying $S$ in \eqref{Raction} with respect to the auxiliary field $\Phi$, we find $\Phi = R$, implying that this action is equivalent to \eqref{fphiRact}. Treating $\Phi$ as an independent scalar field, we may express the action as
\begin{equation}
 \begin{aligned}
S = \int  d^4 x\,  \sqrt{-g}  \, \bigg\{\!
-\frac{1 }{2}   F(\varphi, \Phi)_{,\Phi}R 
&+ \frac{k_{AB}}{2} g^{\mu\nu }(\nabla_\mu \varphi^A) (\nabla_\nu \varphi^B) 
 \\
 &- \frac{ F(\varphi,\Phi) }{2}   +\frac{ 1}{2} F(\varphi,\Phi)_{,\Phi}\Phi - V(\varphi) 
 \bigg\}\,,
\end{aligned}
\end{equation}
from which we may read off the new model functions in terms of $\varphi^M = (\varphi^A, \Phi )$:
\begin{equation}
\label{fRparams}
 \begin{aligned}
f (\varphi, \Phi)&= F(\varphi, \Phi)_{,\Phi}\; ,\\
k_{MN} (\varphi, \Phi)& = \text{diag}(0,k_{AB})\; ,\\
V(\varphi, \Phi)&= V(\varphi) 
 +\frac{F(\varphi,\Phi) - F(\varphi,\Phi)_{,\Phi}\Phi}{2}\; ,
\end{aligned}
\end{equation}
where the indices $M$ and $N$ run over $1$ to $n$ as well as $0$, which corresponds to the scalar curvature $\Phi = R$. These new model functions have the same transformation properties as the ones in \eqref{transrules} and may be used to calculate the potential slow-roll parameters for the theory, leading to analytic predictions in terms of $\Phi$ and $\varphi$. We note that even in the presence of a single scalar degree of freedom $ \varphi$, isocurvature modes may still be generated as the perturbations~$\delta \Phi$ and~$\delta \varphi$ are independent. For $F(\varphi,R)$ models with only one scalar degree of freedom, it is possible to write down the only independent component of the Riemann curvature tensor. For instance, for large values of $\varphi$, we assume that
$F(\varphi, R)$ may be expanded as
 \begin{align}
   \label{eq:Fexample}
  F(\varphi, R) \ = \ a_0(R)\ +\ \frac{a_2 (R)}{ \varphi^2}\ +\ \,{\cal O}(1/\varphi^4)\ .
\end{align}
For this generic class of $F(\varphi, R)$ theories, 
the field-space Ricci scalar $S$ is found to be
  \begin{align}
S\ = \ -2a_2(R)\;.
\end{align}
If the coefficient $a_2(R)$ in the expansion~\eqref{eq:Fexample} of $F(\varphi,R)$ is positive, then the field-space Ricci scalar $S$ is negative, giving rise to a negative contribution to $\bar\eta_{ss}$ [cf.~\eqref{eta2}]. As a consequence, if the frame-invariant potential $U$ happens to be concave, a positive $a_2(R)$ will create further amplification of entropy perturbations, which in turn can 
affect the adiabatic perturbations and so produce relevant effects on the observable CMB power spectrum.

\section{The Vilkovisky--De Witt Formalism for Conformal Transformations}
\label{radcorr}

We have developed a covariant formalism applicable to multifield theories of inflation in the Born approximation. However, the question of whether frame covariance, which has led to frame-invariant classical predictions, may be extended beyond the tree level remains open. In most models of monomial inflation, the radiative corrections generated from the quantum fluctuations of the scalar fields to the matter sector are argued to be suppressed. However, as we are entering an era of precision cosmology, it is desirable to calculate their phenomenological impact. In this respect, it is natural to extend the concept of frame covariance to incorporate radiative corrections. In this section, we outline the Vilkovisky--De Witt formalism \cite{Vilkovisky:1984st,Rebhan:1986wp,DeWitt} and extend it such that it takes into account conformal transformations, thereby leading to an extension of frame covariance beyond the tree-level.

To simplify our analysis, we assume that the scalar fields do not couple to other matter fields and that the radiative corrections due to the quantum fluctuations of the metric $g_{\mu\nu}$ are negligible compared to the radiative corrections due to the scalar fields. Our starting point is the \emph{effective action}~$\Gamma[\varphi] \equiv \Gamma[g_{\mu\nu},\varphi, f(\varphi), k_{AB}(\varphi), V(\varphi)]$.  In order to derive an expression for the effective action, we first write down the \emph{partition functional} $Z[J]$ corresponding to the classical action $S[\varphi]$,
\begin{align}
\label{zdef}
Z[J]\ &=\ \int [\mathcal{D}\phi]\, \mathcal{M} [\phi]\, \exp \left[\frac{i}{\hbar} \Big( S[\phi] +  J_a \phi^a \Big)\right].
\end{align}
In~\eqref{zdef},  $J^a$ are the external sources, $\phi^a$ denote the quantum fields, whereas $\varphi^a$ is reserved to denote below the mean fields. In addition, $[\mathcal{D}\phi]$ stands for integration over all possible paths and $\mathcal{M}[\phi]$ is the associated functional measure of integration which may be determined by the Hamiltonian approach to quantizing the theory.  In \eqref{zdef} and the following, we use the Einstein--De Witt notation, in which $\phi^a \equiv \phi^A(x_A)$ (and similarly for $\varphi^a$) and repeated indices are both summed and integrated over, e.g.
\begin{align}
J_a \varphi^a \ \equiv\ \int d^4 x_A \, \sqrt{-g }\; J_A(x_A) \varphi^A (x_A)\;.
\end{align}
We then consider the generating functional of all connected diagrams $W[J]$, which is defined with the help of $Z[J]$ as follows:
\begin{align}
W[J]\ \equiv\ -i\hbar \ln Z[J]\;.
\end{align}
The corresponding mean field $\varphi^a$ in the presence of external sources $J_a$ is obtained via a functional derivative
\begin{align}\label{phiJ}
 \varphi^a (J)\ \equiv\ \braket{\varphi^a } =  \frac{\delta W[J]}{\delta J_a}\;.
\end{align}
Finally, the effective action in terms of the mean fields $\varphi$ may be written as a Legendre transform of $W[J]$,
\begin{align}
   \label{legtrans}
\Gamma[  \varphi] \ =\ W[J(\varphi)] -  J_a {  \varphi}^a\;.
\end{align}
Note that the source field $J_a(\varphi)$ is given by
\begin{align}\label{gammaJ}
\Gamma_{,a}\ \equiv\ \frac{\delta \Gamma[\varphi]}{\delta\varphi^a}\ =\ -\,J_a\;,
\end{align}
where $\Gamma_{,a}$ stands for the functional derivative with respect to the field~$\varphi^a\equiv \varphi^A(x_A)$ throughout this section.

Writing $W[J]$ and $Z[J]$ in terms of $\Gamma[\varphi]$ through~\eqref{legtrans} and~\eqref{gammaJ}, we may derive the following integro-differential equation:
\begin{align}\label{effactorig}
\exp\left(\frac{ i}{\hbar} \Gamma [ \varphi]     \right) \ =\  \int  [ \mathcal{D}\phi]\, \mathcal{M} [\phi]\, \exp \left\{\frac{i}{\hbar} \Big[ S[\phi] +   \Gamma_{,a} \big(\varphi^a-\phi^a\big) \Big]\right\}\,.
\end{align}
It is obvious that this equation is not covariant due to the presence of the non-covariant term~$\varphi^a -\phi^a$, simply because $\varphi^a$ does not behave as a frame-covariant tensor. To remedy this problem, De Witt imposed the requirement that the action should reduce to its traditional form when expressed in the functional equivalent of normal coordinates. This constraint motivates the replacement of $\varphi^a -\phi^a$ by some two-point covariant function $\Sigma^a (\varphi,\phi )$~\cite{DeWitt,Rebhan:1986wp} to be determined below. In this framework, the quantum fields $\phi^a(x)$ take on the role of coordinates parameterizing the \emph{configuration space}. Although De Witt's original formulation did not consider conformal transformations, it would be preferable to extend his idea to include them. To do so, we define frame-covariant quantities in configuration space through a straightforward extension of the transformation property~\eqref{covdef},
\begin{align}
\widetilde  X^{\widetilde a_1 \widetilde a_2 \ldots \widetilde a_p}_{\widetilde b_1 \widetilde b_2 \ldots \widetilde b_q} \ &=\ \Omega^{-d_X} \big( K^{\widetilde a_1}_{\ a_1} K^{\widetilde a_2}_{\ a_2} \ldots K^{\widetilde a_p}_{\ a_p}\big)  \ X^{  a_1   a_2 \ldots a_p}_{b_1 b_2 \ldots b_q} \ \big(K^{ b_1}_{\ \widetilde b_1} K^{ b_2} _{\ \widetilde b_2}\ldots  K^{ b_q} _{\ \widetilde b_q}\big)\; ,
\end{align}
where the scaling dimension $d_X$ and the conformal weight $w_X$ are related through the expression $d_X = w_X + p -q$, which remains unchanged from \eqref{scaldimrel}. The configuration space Jacobian~$K^{\tilde a}_{\ b}$ is given by
\begin{align}
  K^{\tilde a}_{\ b} \ \equiv\ \Omega^{-1} K^{\widetilde A}_{\ B}\, \delta(x_{\widetilde A} -x_B)\; .
\end{align}
We have thus extended the notion of frame covariance to configuration space.   We note that the configuration space inherits all frame-covariant properties of the field space of the classical theory. Therefore, our approach follows along the lines of the covariantization procedure discussed in Section \ref{sec:frametrans}. 

Given that the configuration space is a manifold, we may now define the associated metric $\mathcal{G}_{ab}$ induced by the field space metric~$G_{AB}$ stated in~\eqref{eq:GAB} as
\begin{align}
   \label{eq:calGAB}
 \mathcal{G}_{ab} \ \equiv\ G_{AB}\: \delta(x_A-x_B)\;.
\end{align}
Following Vilkovisky's approach, we may introduce a field-covariant functional derivative~$\nabla_a$ that respects field reparametrizations \cite{Vilkovisky:1984st} with analogous transformation properties to $\nabla_A$ as given by \eqref{covder}, i.e.
\begin{align}
    \label{covfuncder}
\nabla_{\tilde c}   {\widetilde X}^{\tilde a_1 \tilde a_2 \ldots a_p}_{\tilde b_1 \tilde b_2 \ldots b_q}\ =\ \Omega^{-(d_X-1)} 
\big(K^{\tilde a_1}_{\  a_1}K^{\tilde a_2}_{\  a_2} \ldots  K^{ a_p} _{\ \widetilde a_p} \big)\,  
 \big(\nabla_{  c}\,  X^{a_1 a_2 \ldots a_p}_{b_1 b_2 \ldots b_q}\big) \,
\big(K^{ b_1}_{\ \tilde b_1}K^{ b_2}_{\ \tilde b_2} \ldots  K^{ b_q} _{\ \widetilde b_q}\big) \; .
\end{align}
Proceeding as in \eqref{confcovfieldder}, we may determine the form of this derivative by first defining the \emph{conformally-covariant functional derivative} as
\begin{align}
\label{funcconfder}
  X^{a_1 a_2\ldots a_p}_{b_1 b_2\ldots b_q; c} \ \equiv\ 
    X^{a_1 a_2\ldots a_p}_{b_1 b_2\ldots b_q,c} - \frac{w_X}{2} \frac{  f_{,c}}{f} X^{a_1 a_2\ldots a_p}_{b_1 b_2\ldots b_q} \;.
\end{align}
Then, we may write the  fully \emph{frame-covariant functional derivative} by extending \eqref{funcconfder} to include reparametrizations as done in~\eqref{fieldcovderdef},
\begin{equation}
\label{covfuncderform}
\begin{aligned}
\nabla_c X^{a_1 a_2\ldots a_p}_{b_1 b_2\ldots b_q}\ \equiv\   X^{a_1 a_2\ldots a_p}_{b_1 b_2\ldots b_q;c} &+ \Gamma^{a_1}_{cd} X^{d a_2\ldots a_p}_{b_1 b_2\ldots b_q}+\cdots + \Gamma^{a_2}_{cd} X^{a_1 a_2\ldots d}_{b_1 b_2\ldots b_q}  
\\
&- \Gamma^{d}_{b_1 c} X^{a_1 a_2\ldots a_p}_{d b_2\ldots b_q} - \cdots -  \Gamma^{d}_{b_1 c} X^{a_1 a_2\ldots a_p}_{b_1 b_2 \ldots d}\; ,
\end{aligned}
\end{equation}
where the conformally-covariant configuration space connection $\Gamma^a_{bc}$ is given by
\begin{align}
\Gamma^{a}_{bc}\  &\equiv\ \frac{\mathcal{G}^{ad}}{2}\, \big(\mathcal{G}_{db;c} + \mathcal{G}_{cd;b}  -  \mathcal{G}_{bc;d}\big)\;.
\end{align}

We are now in the position to write a fully frame-covariant form for the effective action. By replacing all quantities with their frame-covariant counterparts, the form of the effective action given in \eqref{effactorig} becomes
\begin{align}\label{effact}
\exp\left(\frac{ i}{\hbar} \Gamma [\varphi]     \right)\ =\  \int [\mathcal{D}\phi]\, \mathcal{M} [\phi]\, \exp \left\{\frac{i}{\hbar} \Big[ S[\phi]\: +\:   (\nabla_a \Gamma) \, (C^{-1}[\varphi])^a_{\ b}\, \sigma^b(\varphi , \phi)   \Big]\right\},
\end{align}
where $\sigma^a$ denotes the tangent tensor to the geodesic connecting $\varphi^a$ to $\phi^a$ and all covariant derivatives are with respect to the mean fields $\varphi^a$. This covariant quantity is defined through
\begin{align}
\label{sigmaprops}
\sigma^b \nabla_b\,\sigma^a\vert_{\varphi=\phi}\ &=\ \sigma^a\; ,
& \sigma^a (\phi,\phi)\ =\ 0\;,
\end{align}
along with the constraint $\det\big(\nabla_a \sigma^b\big) \neq 0$  at $\varphi^a = \phi^a$. The tensor $ (C^{-1}[\varphi])^a_{\ b}$ is found to satisfy the constraint equation 
\begin{align}\label{cdef}
C^a_{\ b}[\varphi] =  \braket{\nabla_b \sigma^a (\varphi,\phi)},
\end{align}
where the brackets $\braket{\ldots}$ denote the functional average calculated with the new form for the action~\eqref{effact}. For a flat configuration space, \eqref{cdef} simplifies to $C^a_{\ b} = \delta^a_b$, leading to the Vilkovisky action. However, in the presence of non-zero curvature, we have \cite{Rebhan:1986wp}
\begin{align}\label{cdef}
\nabla_b \sigma^a(\varphi,\phi) \; = \;  \delta^a_b \;- \; \frac{1}{3} R^a_{\ ibj} \; \sigma^i(\varphi,\phi)  \; \sigma^j(\varphi,\phi) \; +\; \ldots ,
\end{align}
where $ R^a_{\ ibj}$ is the configuration space Riemann tensor.
In this case, we must use the more general Vilkovisky--De Witt action given in \eqref{effact}, where $\Sigma^a (\varphi,\phi ) = (C^{-1}[\varphi])^a_{\ b}\, \sigma^b(\varphi , \phi)$. 
A~unique solution to \eqref{sigmaprops} with the appropriate boundary conditions may be represented by the following non-linear series in powers of $\varphi^a - \phi^a$:
\begin{align}
   \label{sigmaexpansion}
 \sigma^a(\varphi,\phi)\ &=\  \big(\varphi^a - \phi^a\big)\: -\: \frac{1}{2}\,\Gamma^a_{bc}(\varphi)\,\big(\varphi^b -\phi^b\big)\big(\varphi^c -\phi^c\big)\: +\: \mathcal{O}\big[(\varphi-\phi)^3\big],
\end{align}
where $\sigma^a$ can be seen to have scaling dimension $d_\sigma = 1$ and no conformal weight.

The form of the effective action~$\Gamma[\varphi]$ given in \eqref{effact} is unique in the sense that it does not depend on the choice of frame. We may solve~\eqref{effact} in a series of $\hbar$ by redefining the quantum fields as $\phi ^a\rightarrow \varphi^a + \hbar^{1/2} \phi^a$. Then, $\Gamma[\varphi]$ may be expressed in powers of $\hbar$ as
\begin{equation}
  \label{eq:Ghbar}
\begin{aligned}
\Gamma[\varphi]\ &=\ \sum_n \hbar^n \Gamma_n\;,
\end{aligned}
\end{equation}
with $\Gamma_0[\varphi] = S[\varphi]$. The next term of expansion~\eqref{eq:Ghbar} is an invariant expression for the one-loop effective action,
\begin{equation}
   \label{effactder}
\begin{aligned}
\Gamma_1[ \varphi ]\ &=\ -\,i\,
\ln \mathcal{M[\varphi]}\: +\: \frac{i}{2}\, \ln \det \Big( \nabla_a \nabla_b\, S[ \varphi] \Big)\,.
\end{aligned}
\end{equation}
We note that $w_S =w_\mathcal{M} = 0$, and so a conformal functional derivative when acting on the action~$\Gamma_1[ \varphi ]$ reduces to a standard functional derivative. We may verify that $\Gamma_1[\varphi]$ is invariant under a frame transformation due to the property~\eqref{covfuncder} of the frame-covariant derivative and the property for $\mathcal{M}[\varphi]$:
 \begin{align}
 \label{transform}
\mathcal{\widetilde M[\widetilde \varphi]}\ =\  \Omega^n \; |\!\det K| \; \mathcal{ M}[\varphi]\; .
\end{align}
where $n$ is the number of scalar fields. The property \eqref{transform} is a consequence of the fact that the path element measure $[\mathcal{D}\phi] \,\mathcal{M}[\phi]$ remains invariant under a frame transformation. The explicit form of the measure $\mathcal{M}[\varphi]$ is given by the determinant of the metric of the configuration space
\begin{align}
\label{measure}
\mathcal{M}[\varphi ]\ \equiv\ \sqrt{\det  \mathcal{G}_{ab}}\; ,
\end{align}
which satisfies \eqref{transform}. Substituting \eqref{measure} in \eqref{effactder}, we arrive at the frame-invariant  one-loop effective action~\cite{Vilkovisky:1984st,Burgess:1987zi,Ellicott:1987ir}
 \begin{align}
   \label{eq:G1eff}
\Gamma_1[ \varphi ]\ &=\ -\,\frac{i}{2}\,\text{tr} \ln \mathcal{G}_{ab}\: +\:  \frac{i}{2}\, \text{tr} \ln  \Big( \nabla_a \nabla_b\, S[ \varphi] \Big)\ =\  \frac{i}{2}\, \text{tr} \ln  \Big( \nabla^a \nabla_b\, S[ \varphi] \Big)\,.
\end{align}

In this section we have outlined a fully frame-covariant formalism that can be used to compute radiative corrections to the observables due to the quantization of the scalar fields. We may compute the corrected model parameters to higher orders in $\hbar$ by iteratively solving~\eqref{effact}. At the one-loop order, we may explicitly calculate the frame-invariant correction~$\Gamma_1[\varphi]$ to the classical action~$\Gamma_0[\varphi] = S[\varphi]$ through \eqref{eq:G1eff}. Application of the approach developed here 
to specific models of inflation will be given in a forthcoming paper~\cite{BKP}.

\section{Conclusions}
\label{conclusion}

We are entering a new era of precision cosmology, and so the development of theoretical methods for computing observable inflationary quantities with increasingly higher accuracy is of utmost importance. In particular, radiative corrections which might be subleading in several models of inflation may become observationally significant not too far in the future. Hence, any formalism that aspires to remain relevant should describe quantum loop effects in a frame-covariant manner. This is a particularly pressing problem, as frame covariance of the effective potential beyond the Born approximation is still an open issue.  With this motivation, we have developed in this paper a fully frame-covariant formalism of inflation for multifield scalar-curvature theories at the classical level, which may be extended to include radiative corrections.

Making use of notions known from differential geometry, we have adopted an approach in which the scalar fields take on the role of generalized coordinates of a manifold, and the equations of motion describe a trajectory within the field space. By perturbatively expanding and quantizing the resulting frame-covariant equations of motion, we showed how manifestly frame-invariant expressions for inflationary observables may be obtained. These include the tensor-to-scalar ratio $r$, the spectral indices $n_{\cal R}$ and $n_T$, their runnings~$\alpha_{\cal R}$ and~$\alpha_T$, the non-Gaussianity parameter~$f_{NL}$, and the isocurvature faction $\beta_{\rm iso}$. We have further examined the effects of the field space curvature on the generation and transfer of isocurvature modes in a frame-covariant way. 

We have seen in Section \ref{quantpert} that the presence of multiple fields gives rise to more degrees of freedom in selecting an inflationary trajectory. Hence, we have introduced in Section~\ref{subsec:stability} a criterion to discriminate fine-tuned trajectories from stable ones that would require less fine-tuning of the boundary conditions. To this end, we have defined the sensitivity parameter $Q$ in \eqref{relsens}. In order to illustrate our approach, we have considered in Section~\ref{specmod} a simple minimal two-field model, as well as a more realistic nonminimal extension inspired by Higgs inflation. We have observed that after selecting nominal values for the model parameters, the normalization of the power spectrum of scalar perturbations can be used in order to select an appropriate boundary condition for the fields at the end of inflation. We also find that, for certain values of the mass and coupling parameters, the calculation of the entropy transfer becomes crucial in selecting the appropriate trajectory. Furthermore, we have briefly discussed how~$F(\varphi,R)$ theories may be incorporated into our formalism by recasting them in terms of a multifield theory through the method of Lagrange multipliers.

Finally, in Section \ref{radcorr}, we have outlined how the Vilkovisky--De Witt formalism may be applied to multifield inflation. We have extended the notion of classical frame covariance  to the configuration field space in the path integral formulation under the assumption that quantum gravity effects can be ignored. By replacing the functional derivatives with  their covariant counterparts, the so-defined effective action is unique and becomes invariant under frame transformations, which includes both conformal transformations and inflaton reparametrizations. This ensures that scalar-curvature theories related by frame trans\-formations are phenomenologically equivalent. We aim to present a detailed application of the Vilkovisky--De Witt formalism that takes into account radiative corrections including quantum gravity effects for certain multifield models in a future publication.

\subsection*{Acknowledgements}

The authors would like to thank Daniel Burns for collaboration during the early stages of this project and Fedor Bezrukov for helpful discussions and comments. AP also wishes to thank the Mainz Institute for Theoretical Physics for their kind hospitality during which this project was initiated, as well as the organizers and participants of the workshop on ``Exploring the Energy Ladder of the Universe'' (30 May 2016 -- 10 June 2016, Johannes Gutenberg University, Mainz, Germany) for the many stimulating discussions.  The work of SK is supported by an STFC PhD studentship.  The work of AP is supported by the Lancaster--Manchester--Sheffield Consortium for Fundamental Physics, under STFC research grant ST/L000520/1.

\end{document}